\documentclass{aa}
\makeatletter
\renewcommand*\aa@pageof{, page \thepage{} of \pageref*{LastPage}}
\makeatother
\usepackage{pdfpages}			
\usepackage{graphicx}
\usepackage{rotating}
\usepackage[varg]{txfonts}
\usepackage{txfonts}
\usepackage{color}
\usepackage{natbib}
\usepackage{amsmath}
\usepackage{hyperref}
\usepackage{longtable}
\usepackage{multicol}
\bibpunct{(}{)}{;}{a}{}{,} 

\def\bpic{$\beta$ Pic}
\def\dsct{$\delta$~Scuti\,}

\def\Msun{$M_{\odot}$}
\def\Lsun{$L_{\odot}$}
\def\Rsun{$R_{\odot}$}

\def\Teff{\ensuremath{T_{\mathrm{eff}}}}
\def\cd{\,d$^{\rm -1}$}
\def\logg{\ensuremath{\log g}}

\def\vsini{\ensuremath{{\upsilon}\sin i}}
\def\kms{$\mathrm{km\,s}^{-1}$}

\def\OmegaK{\ensuremath{\Omega_{\mathrm{K}}}}
\def\OmegaKmod{\ensuremath{\Omega_{\mathrm{K,\,mod}}}}
\def\sigmaF{\ensuremath{\sigma_{\mathrm{F}}}}

\begin{document}

   \title{Revisiting the pulsational characteristics of the exoplanet host star $\beta$ Pictoris\thanks{Based on data collected by the BRITE Constellation satellite mission, designed, built, launched, operated and supported by the Austrian Research Promotion Agency (FFG), the University of Vienna, the Technical University of Graz, the University of Innsbruck, the Canadian Space Agency (CSA), the University of Toronto Institute for Aerospace Studies (UTIAS), the Foundation for Polish Science \& Technology (FNiTP MNiSW), and National Science Centre (NCN).},\thanks{Based on observations made with ESO Telescopes at the La Silla Paranal Observatory under programme ID 094.D-0274A},\thanks{Light curve data are available at the CDS via anonymous ftp to cdsarc.u-strasbg.fr (130.79.128.5)}}
   \titlerunning{Pulsational characteristics of \bpic}

   \author{K. Zwintz
          \inst{1} \and
          D. R. Reese \inst{2} \and 
          C. Neiner \inst{2} \and 
          A. Pigulski\inst{3} \and 
          R. Kuschnig\inst{4} \and 
          M. M\"ullner\inst{1} \and
          S. Zieba\inst{1} \and 
          L. Abe\inst{5} \and
          T. Guillot\inst{5} \and 
          G. Handler\inst{6} \and
          M. Kenworthy\inst{7} \and 
          R. Stuik\inst{7} \and
          A. F. J. Moffat\inst{8} \and
          A. Popowicz\inst{9} \and
          S. M. Rucinski\inst{10} \and 
          G. A. Wade\inst{11} \and 
          W. W. Weiss\inst{12} \and
          J. I. Bailey III\inst{13} \and 
          S. Crawford\inst{14,15} \and 
		  M. Ireland\inst{16} \and 
          B. Lomberg\inst{15,17} \and 
          E. E. Mamajek\inst{18,19} \and 
          S. N. Mellon\inst{19} \and 
          G. J. Talens\inst{7,20}
                  }
\authorrunning{Zwintz, K., et al.}
   \institute{Institut f\"ur Astro- und Teilchenphysik, Universit\"at Innsbruck, Technikerstra{\ss}e 25, A-6020 Innsbruck\\
              \email{konstanze.zwintz@uibk.ac.at}
   \and 
LESIA, Observatoire de Paris, Universit\'e PSL, CNRS, Sorbonne Universit\'e, Univ. Paris Diderot, Sorbonne Paris Cit\'e, 5 place Jules Janssen, 92195 Meudon, France
         \and
Instytut Astronomiczny, Uniwersytet Wroclawski, Kopernika 11, 51-622 Wroclaw, Poland
		\and 
Institut f\"ur Kommunikationsnetze und Satellitenkommunikation, Technical University Graz, Inffeldgasse 12, A-8010 Graz, Austria 
     \and 
Universit\'e C\^{o}te d’Azur, Observatoire de la C\^{o}te d’Azur, CNRS, Laboratoire Lagrange, CS 34229, 06304 Nice Cedex 4, France
     \and 
Nicolaus Copernicus Astronomical Center, ul. Bartycka 18, 00-716 Warsaw, Poland     \and 
Leiden Observatory, Leiden University, P.O. Box 9513, 2300 RA Leiden, The Netherlands
\and 
D\'epartement de physique, Universit\'e de Montr\'eal, CP 6128, Succursale Centre-Ville, Montr\'eal, Qu\'ebec H3C 3J7, Canada; Centre de Recherche en Astrophysique du Qu\'ebec (CRAQ), Montr\'eal, Qu\'ebec H3C 3J7, Canada
\and 
Institute of Automatic Control, Silesian University of Technology, Akademicka 16, Gliwice, Poland
\and 
Department of Astronomy and Astrophysics, University of Toronto, 50 St. George Street, Toronto, Ontario, M5S 3H4, Canada
\and 
Department of Physics and Space Science, Royal Military College of Canada, PO Box 17000, Stn Forces, Kingston, K7K 7B4 Ontario, Canada
\and 
Universit\"at Wien, Institut f\"ur Astrophysik, T\"urkenschanzstrasse 17, A-1180 Wien \and 
Department of Physics, University of California at Santa Barbara, Santa Barbara, CA 93111 USA \and
Space Telescope Science Institute, 3700 San Martin Drive, Baltimore,
MD 21218, USA \and 
South African Astronomical Observatory, Observatory Rd, Observatory
Cape Town, 7700 Cape Town, South Africa \and
Research School of Astronomy and Astrophysics, Australian National University, Canberra, ACT 2611, Australia \and 
Department of Astronomy, University of Cape Town, Rondebosch, 7700 Cape Town, South Africa \and 
Jet Propulsion Laboratory, California Institute of Technology, M/S 321-100, 4800 Oak Grove Drive, Pasadena, CA 91109, USA \and
Department of Physics \& Astronomy, University of Rochester, 500 Wilson Blvd., Rochester, NY 14627, USA \and 
Institut de Recherche sur les Exoplan\`etes, D\'epartement de Physique, Universit\'e de Montr\'eal, Montr\'eal, QC H3C 3J7, Canada
}

   \date{Received; accepted}

 
  \abstract
   {Exoplanet properties crucially depend on their host stars' parameters: more accurate stellar parameters yield more accurate exoplanet characteristics. 
   In case the exoplanet host star shows pulsations, asteroseismology can be used for an improved description of the stellar parameters.
   }
   {We aim to revisit the pulsational properties of \bpic\ and identify its pulsation modes from normalised amplitudes in five different passbands. We also investigate the potential presence of a magnetic field.}
   {We conduct a frequency analysis using three seasons of BRITE-Constellation observations in the BRITE 
   filters, the $\sim$620-day long bRing light curve and the nearly 8-year long SMEI photometric time series. We calculate normalised amplitudes using all passbands  including previously published values obtained from ASTEP observations. We investigate the magnetic properties of \bpic\ using spectropolarimetric observations conducted with the HARPSpol instrument.  Using 2D rotating models, we fit the  normalised amplitudes and frequencies through Monte Carlo Markov Chains.}
   {We identify 15 pulsation frequencies in the range from 34 to 55\,\cd, where two -- F13 at 53.6917\,\cd\ and F11 at 50.4921\,\cd\ -- display clear amplitude variability. We use the normalised amplitudes in up to five passbands to identify the modes as three $\ell = 1$, six $\ell = 2$ and six $\ell = 3$ modes. \bpic\ is shown to be non-magnetic with an upper limit of the possible undetected dipolar field of 300\,Gauss. }
   {Multiple fits to the frequencies and normalised amplitudes are obtained including one with a near equator-on inclination for \bpic, which corresponds to our expectations based on the orbital inclination of \bpic\,b and the orientation of the circumstellar disk. This solution leads to a rotation rate of $27$\% of the Keplerian break-up velocity, a radius of 1.497$\pm$0.025\,\Rsun, and a mass of 1.797$\pm$0.035\,\Msun. The $\sim$2\% errors in radius and mass do not account for uncertainties in the models and a potentially erroneous mode-identification.}

   \keywords{Asteroseismology --
                Stars:individual:$\beta$ Pictoris --
                Stars: interiors -- Stars:variables:delta Scuti -- Stars: magnetic field
               }

   \maketitle
%

\section{Introduction}

The description of the formation, structure and evolution of young stars is one of the big challenges in stellar astrophysics. Early stellar evolution plays a crucial role in our understanding of the formation and evolution of exoplanets, whose properties depend on the accuracy of the inferred stellar parameters. 
In many cases, stellar activity (star spots, magnetic fields, pulsations, circumstellar material etc.) complicates the reliable determination of the physical properties of stars from the combination of spectroscopic observations and theoretical models and affects the investigation of the exoplanet properties. However, the presence of pulsations might also be beneficial because asteroseismic methods can be used to constrain the interior structure of exoplanet host stars. With it, more reliable results for the fundamental stellar parameters can be inferred which in turn affects the precision of the derived exoplanet properties.
In this context, the $\beta$ Pictoris system is quite an interesting object.

$\beta$ Pictoris (HD\,39060, spectral type A6\,V) is a bright star ({\it V} = 3.86\,mag) located relatively close to us at 19.76 pc distance \citep[calculated using a parallax of 50.623 $\pm$ 0.334 mas as given in the Gaia DR2 catalog;][]{gaia-dr2}. It is a member of the \bpic\ moving group \citep{mamajek2014}. The star \bpic\ is surrounded by a gas and dust debris disk which is seen nearly edge-on; its outer extent varies from 1450 to 1835 Astronomical Units (AU) \citep{larwood2001}. The observed warp of its inner disk suggested the presence of a planet in the system. Using observations with the VLT at Paranal and the NACO camera, \citet{lagrange2009,lagrange2010} directly imaged the giant gas planet, \bpic\ b, for the first time. Although its orbital inclination is close to equator-on, i.e. 88.81$\pm 0.12^{\circ}$ as seen from Earth, it is sufficiently inclined that \bpic\ b does not transit its host star \citep{wang2016}. The reported four distinct belts around the star kinematically indicate the presence of other planets \citep{wahhaj2003}. So far, no additional planet was detected \citep[e.g.,][]{lous2018}.

The age of the \bpic\ system was investigated by several authors using many different methods \citep[for a review see][]{mamajek2014}. Currently, there seems to be agreement that the age of the \bpic\ system is $\sim$23 Myr \citep{mamajek2014}. From both spectroscopic observations and the derived age, it is evident that the star is in its early main sequence stage of evolution \citep[e.g.,][]{zwintz2014}.

\dsct pulsations in \bpic\ were first discovered 2003 through ground-based photometric time series \citep{koen2003a} where three low-amplitude modes were identified. 
Subsequent spectroscopic time series observations obtained with the 1.9\,m telescope at the South African Astronomical Observatory (SAAO) revealed 18 \dsct pulsation frequencies in the range from 24 to 71\cd\ from line profile variations \citep{koen2003b}.
Recently, \citet{mekarnia17} reported 31 pulsation frequencies between 34.76 and 75.68\cd derived from photometric time series obtained between March and September 2017 using the 40-cm ASTEP telescope at Concordia Station in Antarctica.

Asteroseismology has been successfully used to study different types of pulsating stars from the pre-main sequence (pre-MS) to the final stages of evolution (e.g., in white dwarfs) in a mass range from $\sim$0.5 to 40\,\Msun\ with effective temperatures between $\sim$3\,000\,K and 100\,000\,K. 
\dsct stars are located in the lower part of the so-called classical instability strip where it intersects with the main sequence. 
They have spectral types from A2 to F2 \citep{rodriguez2001} and masses between 1.5 and 4\,\Msun\ \citep[e.g.,][]{aerts2010} and lie in the effective temperature range between 6300 and 8600\,K \citep{uytterhoeven2011}. 
\dsct pulsations can be found in the pre-main sequence, main sequence and post-main sequence evolutionary stages and are driven by the heat-engine ($\kappa$-) mechanism acting in the second helium ionization zone \citep[e.g.,][]{aerts2010}. The pulsation modes are radial and non-radial pressure (p) modes with periods in the range from $\sim$18 minutes to 0.3 days. 

Despite the fact that \dsct stars were one of the first types of stars discovered to pulsate, little progress has been made in interpreting their pulsation spectra due to their complexity. \dsct stars are mostly multiperiodic and can show very rich pulsation frequency spectra \citep[e.g.,][]{poretti2009}. As they are moderate to fast rotators \citep{breger2000}, the influence of rotation on the pulsation frequencies cannot be neglected in theoretical models.  One of the first effects of rotation is to split the frequencies of modes with the same $n$ and $\ell$ values but different $m$ values, thus removing their degeneracy.  At slow rotation rates for uniform rotation profiles, these rotationally split modes should be observed as multiplets with nearly equidistant frequency spacings, but in fact there are only a few cases known where such clear rotational splittings were found \citep[e.g.,][]{kurtz2014}.  At faster rotation rates, the multiplets become non-equidistant as higher-order effects of rotation intervene \citep[e.g.,][]{saio1981, espinosa2004}, and they start to overlap making the spectrum more complicated to interpret \citep{reese2006}.  Eventually, the frequency spectrum of acoustic modes takes on a new structure composed of overlapping independently-organized subspectra associated with different classes of modes \citep{lignieres2008, lignieres2009}.

Furthermore, the pulsation amplitudes of \dsct stars can be variable for different reasons. A good overview of this topic can be found in \citet{bowman2016}, where the authors explain the two intrinsic causes -- (i) beating of a pair of close, unresolved pulsation frequencies and (ii) non-linearity and coupling of modes -- and one extrinsic cause, i.e. binary or multiple systems, of amplitude modulations. 
In the same study, the authors reveal that 61.3\% of their sample of 983 \dsct stars include at least one pulsation mode showing amplitude modulation which illustrates that this is a quite common effect in this group of stars. 

Another complicating factor in the interpretation of \dsct type pulsations can be the presence of a magnetic field. \dsct and rapidly oscillating, chemically peculiar A (roAp) stars are located in the same region of the HR diagram. The roAp stars \citep{kurtz2006} show pulsation periods between 6 and 24 minutes \citep[e.g.,][]{smalley2015} and possess global magnetic fields, inhomogeneous surface distributions of some chemical elements and strong overabundances, including the rare-earth elements \citep{ryabchikova2004}. While the presence of sometimes quite strong magnetic fields are typical for roAp stars, for only very few \dsct stars have magnetic fields been measured \citep[e.g.,][]{kurtz2008,neiner-lampens2015} or were suggested due to the presence of the aforementioned rare-earth anomaly \citep{escorza2016}. 

In the present study we use photometric time series obtained by the BRITE-Constellation satellites in two filters in three observing seasons in combination with photometric time series observed with the SMEI satellite \citep{jackson2004,howard2013}, the bRing instrument \citep{stuik2017} and the previously published results by \citet{mekarnia17} based on data from the ASTEP telescope in Antarctica \citep{abe2013,guillot2015,mekarnia2016} to constrain the pulsational properties of \bpic, identify its pulsation modes from the multiband photometry and investigate the presence of amplitude modulation. Additionally, we use HARPSpol \citep{piskunov11} spectropolarimetric data to investigate the presence of a magnetic field.

\section{Observations}

\begin{figure*}
\begin{center}
\includegraphics[width=0.99\textwidth]{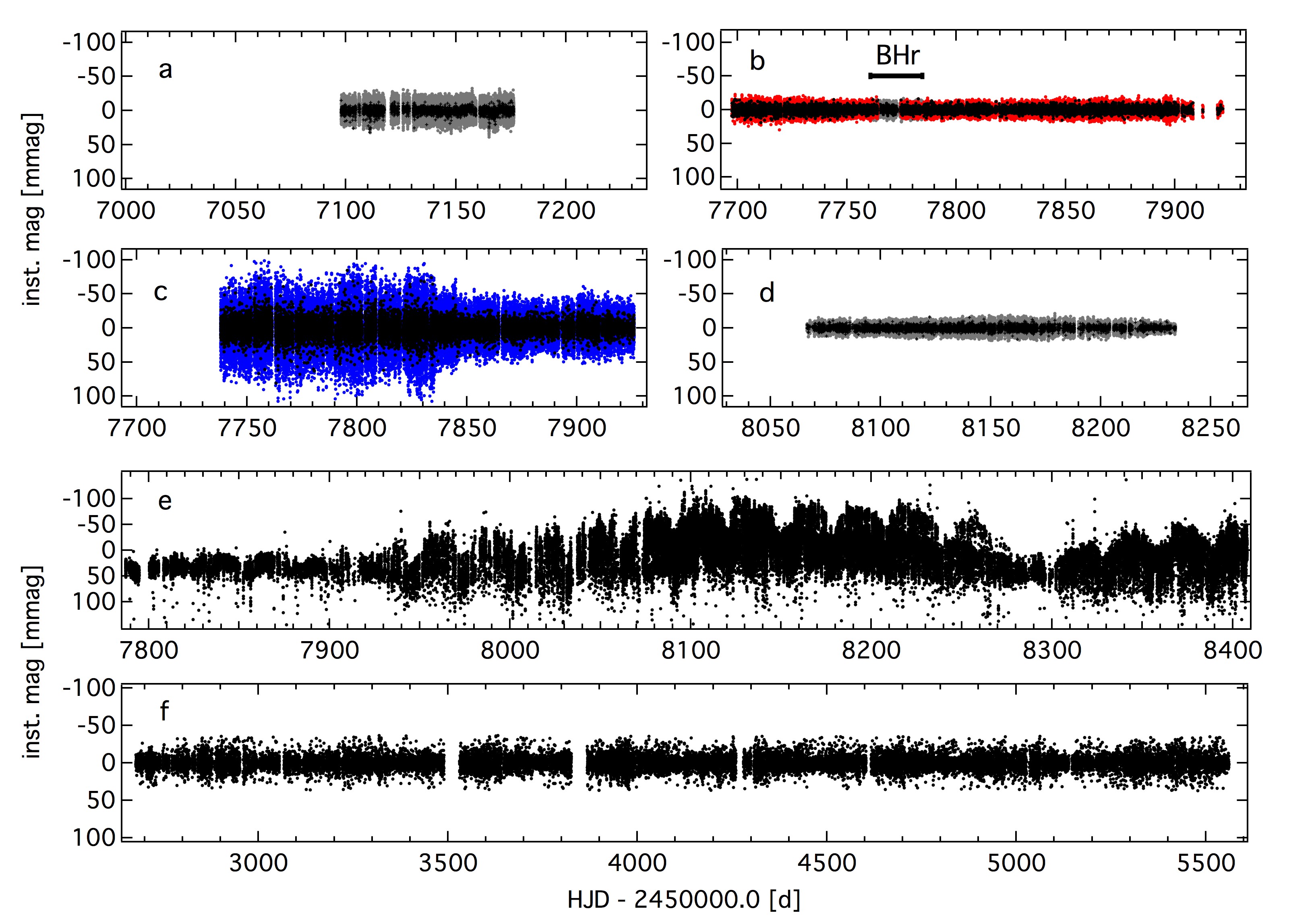}
\caption{Final reduced light curves of \bpic\, from BRITE-Constellation, bRing and SMEI: BHr data 2015 (panel a, grey points), combined BTr (panel b, red points) and BHr (panel b, grey points) data 2016/17, BLb data 2016/17 (panel c, blue points), BHr data 2017/18 (panel d, grey points) where black dots show binning of the light curves in three-minute intervals; the bRing and SMEI time series are shown in panels e and f.}
\label{all-lcs}
\end{center}
\end{figure*}

\subsection{BRITE-Constellation}

BRITE-Constellation\footnote{http://www.brite-constellation.at} consists of five 20-cm cube nanosatellites each carrying a 3-cm telescope and feeding an uncooled CCD \citep{weiss2014}. 
Three BRITE satellites -- i.e., BRITE-Toronto (BTr), Uni-BRITE (UBr) and BRITE-Heweliusz (BHr) -- carry a custom-defined red filter (550 -- 700\,nm), and two satellites -- i.e., BRITE-Austria (BAb) and BRITE-Lem (BLb) -- a custom-defined blue filter (390 -- 460\,nm). More details on the detectors, pre-launch and in-orbit tests are described by \citet{pablo2016}. \citet{popowicz2017} describe the pipeline that processes the observed images yielding the instrumental magnitudes which are delivered to the users.

BRITE-Constellation observes large fields with typically 15 to 20 stars brighter than $V=6$ mag including at least three targets brighter than $V=3$ mag. Each field is observed at least 15 minutes per each $\sim$100-minute orbit for up to half a year \citep{weiss2014}. 

BRITE-Constellation first obtained observations of \bpic\ from 16 March - 2 June 2015 (BRITE Run ID: 08-VelPic-I-2015), yielding a total time base of 78.323 days using BHr in stare mode \citep{popowicz2017}. Hence, the corresponding frequency resolution, 1/T, is 0.013\,\cd.
After the success of this first observing season, a longer observing run was conducted using BTr from 4 November 2016 to 17 June 2017 for a total of 224.573 days and BLb from 15 December 2016 to 21 June 2017 for 187.923 days (BRITE Run ID: 23-VelPic-II-2016). BHr was used from 7 January 2017 to 30 January 2017 for 24 days to cover a gap in the BTr observations. 
Recently, the BRITE-Constellation observations of the third season for \bpic\ were completed. The red BHr satellite obtained time series of \bpic\ between 9 November 2017 to 25 April 2018 for 167.335 days (BRITE Run ID: 33-VelPicIII-2017).
The 2016/2017 and 2017/2018 observations were made using the chopping mode where the position of the target star within the CCD plane is constantly alternated between two positions about 20 pixels apart on the CCD \citep{popowicz2017}. 
An overview of the BRITE-Constellation observations is given in Table \ref{tab:obs}. Publicly available BRITE-Constellation data can be retrieved from the BRITE Public Data Archive (https://brite.camk.edu.pl/pub/index.html).

\begin{table*}[htb]
\caption{Properties of the multi-colour observations for $\beta$ Pictoris.}
\label{tab:obs}
\begin{center}
\begin{tabular}{llccrcccc}
\hline
 \multicolumn{1}{c}{Obs ID} & \multicolumn{1}{c}{wavelength} &\multicolumn{1}{c}{obs$_{\rm start}$}  & \multicolumn{1}{c}{obs$_{\rm end}$} & \multicolumn{1}{c}{time base} & \multicolumn{1}{c}{1/T} & \multicolumn{1}{c}{N} & \multicolumn{1}{c}{res. noise} & \multicolumn{1}{c}{$f_{\rm Nyquist}$} \\
\multicolumn{1}{c}{ } & \multicolumn{1}{c}{[nm]}  & \multicolumn{1}{c}{[d] }   & \multicolumn{1}{c}{[d] }   & \multicolumn{1}{c}{[d]} & \multicolumn{1}{c}{[\cd]} & \multicolumn{1}{c}{\#} & \multicolumn{1}{c}{[ppm]} &  \multicolumn{1}{c}{ [\cd]} \\
\hline
 BHr & 550 -- 700 & 16 March 2015 & 2 June 2015 & 78.323 & 0.013 & 44236 & 100 & 4181 \\
 BLb & 390 -- 460  & 15 Dec. 2016 & 21 June 2017 & 187.923 & 0.005 & 74306 & 170 & 2130 \\
 BTr & 550 -- 700 &  4 Nov. 2016 & 17 June 2017 & 224.573 & 0.004 & 53620 & 47 & 2089 \\
 BHr & 550 -- 700  & 7 Jan. 2017 & 30 Jan. 2017 & 23.722 & 0.042 & 13958 & $\ast$ & 2130 \\
 {\it BTr + BHr} & 550 -- 700  & {\it 4 Nov. 2016} & {\it 17 June 2017} & {\it 224.573} & {\it 0.004} & {\it 67578} & {\it 40} & {\it 2146} \\
  BHr & 550 -- 700  & 9 Nov. 2017 & 25 April 2018 & 167.335 & 0.006 & 53262 & 43 & 2127 \\
\hline
SMEI &  450 -- 950  & 6 Feb. 2003  & 30 Dec. 2010  & 2884.609  & 0.0003   &  28623   & 92  & 7 \\
bRing   & 463 -- 639 & 2 Feb. 2017   & 16 Oct. 2018   & 620.249  &  0.002  & 68126   &  110  & 135 \\
\hline
\end{tabular}
\end{center}
\tablefoot{Instruments which provided photometric time series (Obs ID), wavelength range (wavelength), corresponding start (obs$_{\rm start}$) and end dates (obs$_{\rm end}$), total time base of the reduced data set (time base, T), Rayleigh frequency resolution (1/T), number of data points (N), residual noise in the amplitude spectrum after prewhitening all frequencies (res. noise) and Nyquist frequency ($f_{\rm Nyquist}$). $\ast$: The January 2017 BHr data set was only used to fill the gap of the BTr data set, hence was not analyzed by itself. The residual noise level (res. noise) is calculated over the complete range relevant for \dsct pulsations from 0 to 100\,\cd.}
\end{table*}

\subsection{bRing}

bRing (which stands for "the $\beta$ Pictoris b Ring project'') consists of two ground-based observatories which monitored \bpic\ photometrically in particular during the expected transit of the Hill sphere of its giant exoplanet \bpic\ b in 2017 -- 2018 \citep{stuik2017}. One bRing instrument is located in the Sutherland observing station of the South African Astronomical Observatory, the second bRing site is at the Siding Spring Observatory in Australia. Both telescopes take observations in the wavelength range from 463 to 639\,nm.
A detailed description of the design, operations and observing strategy of bRing is provided by \citet{stuik2017}. Both bRing instruments will continue the observations of \bpic\ for as long as possible in the future.

The data set used here was taken between 2 February 2017 and 16 October 2018 with a cadence of $\sim$5 minutes using a combination of the data obtained by both bRing instruments.

\subsection{Solar Mass Ejection Imager (SMEI)}

The Solar Mass Ejection Imager \citep[SMEI;][]{eyles2003,jackson2004,howard2013} was launched as a secondary payload onboard the Coriolis spacecraft in January 2003. Its main purpose was to monitor and predict space weather in the inner solar system. The orbital period of SMEI is $\sim$101.5 minutes \citep[e.g.,][]{eyles2003}. The mission was terminated due to budgetary reasons in September 2011. 
However, the SMEI images have also been shown to yield high quality, long duration stellar photometry. SMEI obtained brightness measurements of nearly the full sky using three cameras with a field-of-view of 3 $\times$ 60 deg$^2$ each. The photometric passband ranges from 450 to 950\,nm. The data rate for the SMEI photometric time series for a single star is one measurement per each $\sim$101.5-minute orbit. Consequently, the Nyquist frequency of the SMEI data lies at 7.08\,\cd.

Stellar time series obtained by SMEI can be extracted from the SMEI website\footnote{http://smei.ucsd.edu/new\_smei/data\&images/stars/timeseries.html}. They have been used several times in the past for a common interpretation with BRITE-Constellation data \citep[e.g.,][]{baade2017,kallinger2017}. In the case of \bpic\, only times and magnitudes were available from the SMEI website with no additional information about the instrumental settings.

The SMEI data for \bpic\ comprise 28623 data points obtained between 6 February 2003 and 30 December 2010 for about eight years in total, which corresponds to a classical Rayleigh frequency resolution of 0.0003\cd\ (see Table \ref{tab:obs}).

\subsection{Spectropolarimetry}

$\beta$\,Pic was observed in conjunction with the BRITE spectropolarimetric survey \citep{neiner14}
with the HARPSpol spectropolarimeter \citep{piskunov11} installed on the ESO 3.6-m telescope in La
Silla (Chile). Observations were acquired on November 7, 2014, and are available through the ESO archive\footnote{archive.eso.org}. A series of seven
consecutive Stokes V sequences were obtained to increase the signal-to-noise
ratio (SNR) of the co-added spectrum while avoiding saturation of the detector.
Each sequence consisted of four sub-exposures of 246\,s, with each sub-exposure in a
different configuration of the polarimeter. This led to a total of almost 2\,h of
exposure for the seven sequences.

The usual bias, flat-field, and ThAr calibrations have been obtained the same
night and applied to the data. The data were reduced using a modified version of
the {\sc REDUCE} software \citep{piskunov2002,makaganiuk2011}. This included
automatic normalisation of the spectra to the intensity continuum level.

\section{Photometric data reduction and frequency analysis}

The frequency analysis of the BRITE, SMEI and bRing photometric time series was performed independently of each other using the software package Period04 \citep{lenz05} that combines Fourier and least-squares algorithms. Frequencies were then prewhitened and considered to be significant if their amplitudes exceeded 3.8 times the local noise level in the amplitude spectrum \citep{breger93,kuschnig97}.
Frequency, amplitude and phase errors are calculated using the formulae given by \citet{montgomery99}.

We verified the analysis using the iterative prewhitening method based on the Lomb-Scargle periodogram that is described by \citet{vanreeth2015aa}.

\subsection{Analysis of the BRITE photometry}

The raw BRITE photometry was  corrected for instrumental effects. The corrections included outlier rejection, and both one- and two-dimensional decorrelations with all available parameters, in accordance with the procedure described by \citet{pigulski2018}.

The data obtained in 2016/17 by BTr and BHr were combined to a single red filter data set. An overview of their properties is given in Table \ref{tab:obs}. Figure \ref{all-lcs} shows the light curves obtained by BHr in 2015 (panel a), by BTr and BHr in 2016/2017 (panel b), by BLb  in 2016/2017 (panel c) and by BHr in 2017/2018 (panel d) to the same Y axis scale. 
For \bpic\, with a $B$ magnitude of 4.03 and a $V$ magnitude of 3.86 significantly less flux is measured through the blue filter than through the red filter which is reflected by a more than a factor of four higher scatter, and hence, also a factor of four higher residual noise level in the frequency analysis (Table \ref{tab:obs}).

\begin{figure*}
\begin{center}
\includegraphics[width=0.9\textwidth]{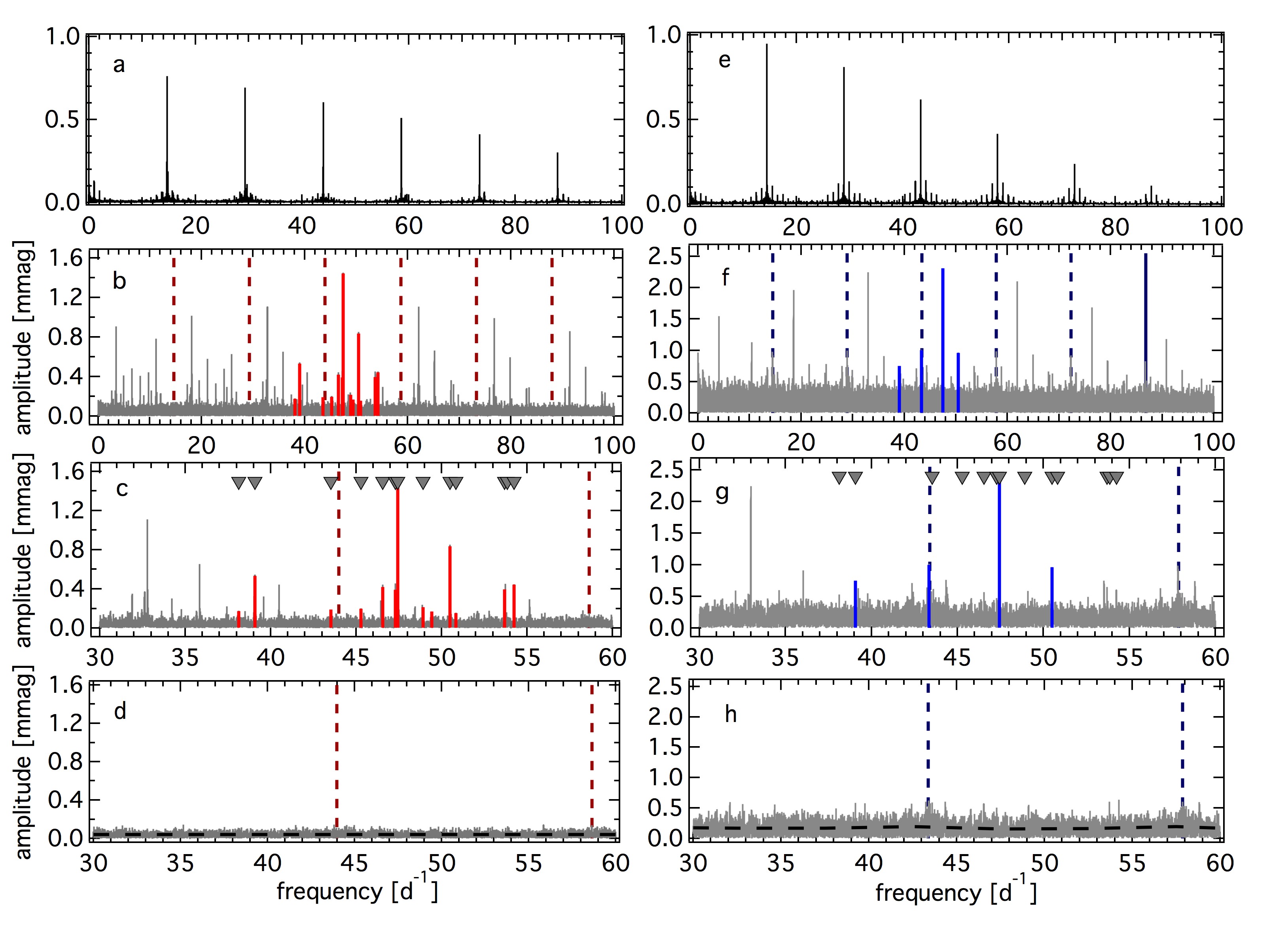}
\caption[]{Frequency analysis of the BRITE 2016/17 data in the red filter (left) and the blue filter (right): spectral window (panels a and e), original amplitude spectrum from 0 to 100\cd\ (panels b and f), zoom into the original amplitude spectrum (panels c and g) and residual amplitude spectrum after prewhitening the corresponding pulsation frequencies (panels d and h) with the residual noise level marked as horizontal dashed lines. The identified pulsation frequencies (as listed in Table \ref{tab:freqs}) are marked in panels b and c as red (for the BRITE red filter) and in panels f and g as blue (for the BRITE blue filter) lines. The triangles mark the frequencies found in the ASTEP data by \citet{mekarnia17}. Vertical dashed lines mark the positions of the respective satellite's orbital frequency (i.e., BTr on the left and BLb on the right) and its multiples.}
\label{2016data}
\end{center}
\end{figure*}

\subsubsection{2015 data}

The frequency analysis of the BHr 2015 data yielded eight intrinsic frequencies with a S/N ratio larger than 3.8.
A comparison to the frequencies reported by \citet{mekarnia17}  shows that there is agreement for six frequencies (F1, F8, F11, F13, F14 and F15; see Table \ref{tab:freqs} and grey triangles in Fig. \ref{BHr2015}). Additionally, we find two frequencies at 32.456\,\cd\ with an amplitude of 0.47$\pm$0.05\,mmag and at 61.367\,\cd\ with an amplitude of 0.49$\pm$0.05\,mmag to be statistically significant with a S/N ratio of 4.66 and 4.92. As these frequencies do not appear in any of our other data sets including those obtained by bRing, SMEI and ASTEP \citep{mekarnia17} , their origin is presently unclear, and, hence, we treat them with caution and discard them from our further analysis.

The residual noise level after prewhitening of the eight frequencies is 100\,ppm. The spectral window function and amplitude spectra are shown in Figure \ref{BHr2015} in the Appendix \ref{appendix.FA}.

\subsubsection{2016/2017 data}

The analysis of the combined 2016/2017 BRITE red filter data set yielded 13 significant pulsation frequencies (Table \ref{tab:freqs}). Only frequency F10 at 49.4161\cd\ was not reported by \citet{mekarnia17}. Although frequency F4 at 43.5268\cd\ is close to three times the BRITE orbital frequency and we would normally have to omit it, we identify it as pulsational because it was already reported by \citet{mekarnia17}.

Due to the significantly higher noise level of the blue filter data set, only four pulsation frequencies were identified from the BLb observations (Table \ref{tab:freqs}). These four frequencies (F3, F4, F8, F11) are also found in the 2015, 2016/2017 and 2017/2018 red filter data sets and have also been reported by \citet{mekarnia17}.

The residual noise level after prewhitening all frequencies is 40\,ppm for the combined red filter and 170\,ppm for the blue filter data.
Figure \ref{2016data} shows the amplitude spectra using the combined BTr and BHr data set (left) and the BLb data (right).

\begin{figure*}[htb]
\begin{center}
\includegraphics[width=0.9\textwidth]{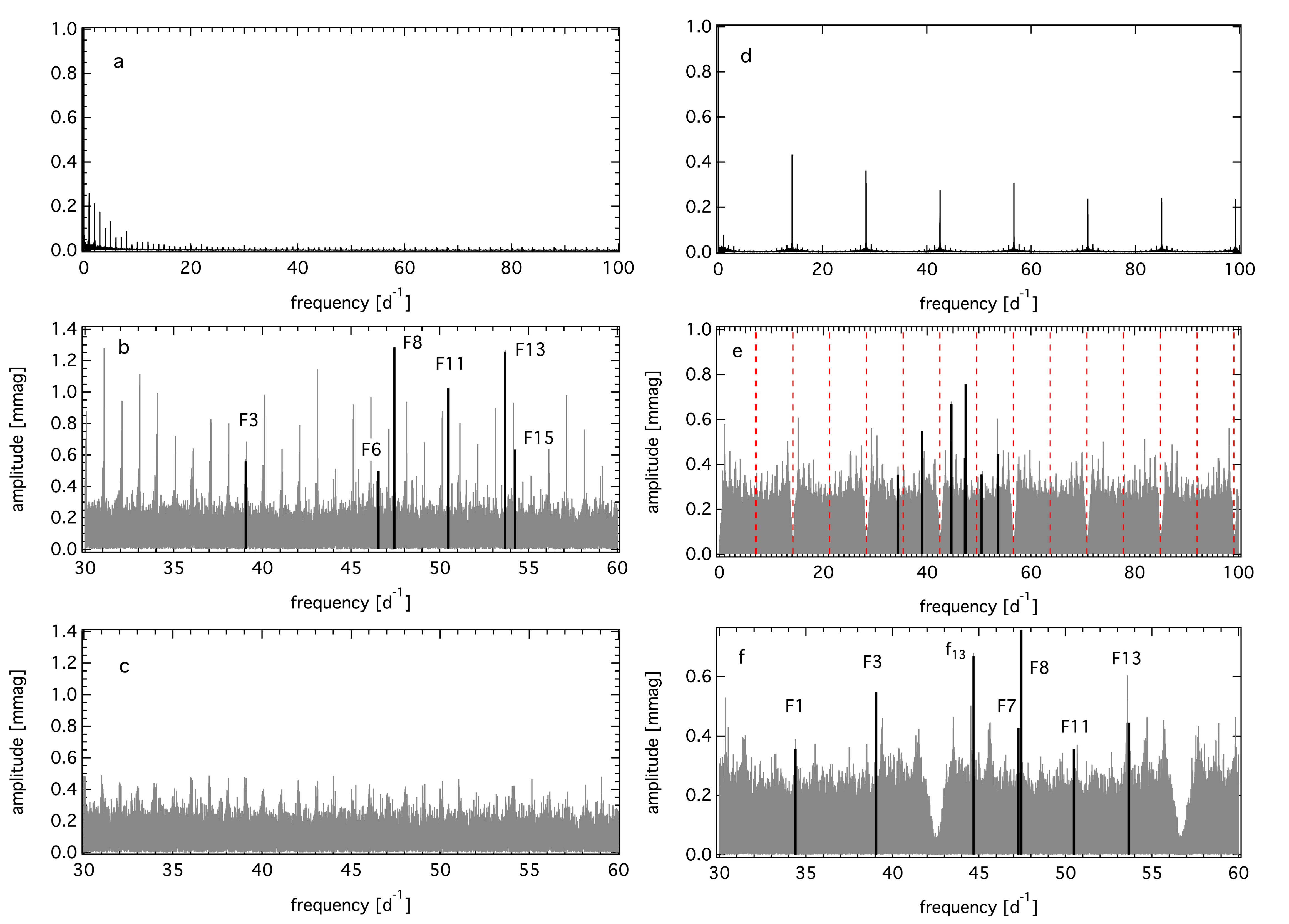}
\caption[]{Frequency analysis using the bRing and SMEI photometric time series: the spectral windows (panels a and d), original amplitude spectrum from 30 to 60\,\cd\ (panel b) for bRing and from 0 to 100\,\cd\ (panel e) for SMEI, the residual bRing amplitude spectrum after subtraction of the six identified pulsation frequencies (panel c) and zoom into the frequency range from 30 to 60\cd\ for SMEI (panel f). The identified pulsation frequencies are marked as thick black lines and labelled according to their numbers in Table\ref{tab:smei_bRing}. Frequency f$_{13}$ was identified in the SMEI data due to its presence in DM17, but is not present in the BRITE data. The red, dashed vertical lines mark the position of the Nyquist frequency for SMEI and its multiples. The occurrence of the orbital frequency of SMEI at 14.16\,\cd\ (i.e., twice $f_{\rm Nyquist}$) and its multiples can be seen as the regular structure in the spectral window (panel d) and in the amplitude spectrum (panels e and f) as the points where the noise decreases steeply to very low levels. }
\label{bring-smei-amp}
\end{center}
\end{figure*}

\subsubsection{2017/2018 data}
With the 2017/18 BHr data set we confirmed seven of the  pulsation frequencies previously identified from the BRITE-Constellation 2016/2017 data and \citet{mekarnia17}. Additionally, two frequencies at 34.085\,\cd\ and 52.960\,\cd\ with amplitudes of 0.27$\pm$0.03\,mmag and 0.20$\pm$0.03\,mmag that were not found in the 2015 and 2016/17 data sets before, are statistically significant with S/N values of 6.34 and 4.77, respectively. As these frequencies do not appear in any of the other observations, we discard them from any further investigation.
The residual noise level after prewhitening all pulsation frequencies is 43\,ppm.

The corresponding amplitude spectrum is shown in Figure \ref{2017data} in the Appendix \ref{appendix.FA}.

\subsubsection{Combined BRITE red filter data}
Combining the three seasons of BRITE-Constellation red filter data yields a total time base of 1135 days corresponding to a Rayleigh frequency resolution, 1/T, of 0.0009\,\cd. All frequencies listed in Table \ref{tab:freqs} can be found in the combined BRITE red filter data set; no additional peaks are statistically significant. The residual noise level calculated from 0 to 100\,\cd\ after prewhitening all significant frequencies lies at 36\,ppm.

Despite the fact that BRITE-Constellation observations are  sensitive in the low frequency domain as was illustrated, e.g., by \citet{baade2018} or \citet{tahina2018}, there is no evidence for the presence of g-modes.

\begin{table*}[htb]
\caption{Pulsation frequencies, amplitudes, phases and signal-to-noise values derived from the BRITE-Constellation data sorted by increasing frequency and comparison to the literature.}
\label{tab:freqs}
\begin{center}
\begin{scriptsize}
\begin{tabular}{rlllcclllrrrl}
\hline
\multicolumn{1}{c}{No.} & \multicolumn{2}{c}{Frequency} & \multicolumn{1}{c}{$A_{R}$ 2015}  & \multicolumn{1}{c}{$A_R$ 2016} & \multicolumn{1}{c}{$A_R$ 2017} & \multicolumn{1}{c}{$\phi_R$ 2015} & \multicolumn{1}{c}{$\phi_R$ 2016} & \multicolumn{1}{c}{$\phi_R$ 2017} & \multicolumn{1}{c}{S/N$_R$ 2015} & \multicolumn{1}{c}{S/N$_R$ 2016}& \multicolumn{1}{c}{S/N$_R$ 2017}& \multicolumn{1}{c}{cross ID} \\
\multicolumn{1}{c}{\#} &  \multicolumn{1}{c}{[\cd]}  &  \multicolumn{1}{c}{[$\mu$Hz]}   & \multicolumn{1}{c}{[mmag]}  & \multicolumn{1}{c}{[mmag]} & \multicolumn{1}{c}{[mmag]} & \multicolumn{1}{c}{ }  & \multicolumn{1}{c}{ } & \multicolumn{1}{c}{ } & \multicolumn{1}{c}{ } & \multicolumn{1}{c}{ } & \multicolumn{1}{c}{ } & \multicolumn{1}{c}{ }  \\
\hline
F1	&	34.4342(9)	&	398.543(11)	&	0.39(5)		&			&	0.27(3)		&	0.61(2)		&			&	0.37(2)		&	3.83	&		&	6.34	&	close to f$_{10}$ in DM17		\\
F2	&	38.1293(4)	&	441.311(4)	&				&	0.17(3)	&				&				&	0.38(2)	&			&			&	4.2	&		& f$_{\rm 11}$ in DM17	\\
F3	&	39.0629(1)	&	452.117(1)	&				&	0.53(3)	&	0.49(3)		&				&	0.059(8)&	0.926(9)&			&	11.1&	10.82	& f$_{\rm 5}$ in DM17		\\
F4	&	43.5268(4)	&	503.783(4)	&				&	0.19(3)	&	0.18(3)		&				&	0.67(2)	&	0.83(2)	&			&	4.3	&	4.1	& f$_{\rm 9}$ in DM17		\\
F5	&	45.2705(4)	&	523.964(4)	&				&	0.20(3)	&				&				&	0.48(2)	&			&			&	4.5	&		& f$_{\rm 12}$ in DM17		\\
F6	&	46.5428(2)	&	538.690(2)	&				&	0.42(3)	&	0.48(3)		&				&	0.22(1)	&	0.515(9)&			&	9.5	&	10.56	& f$_{\rm 6}$ in DM17		\\
F7	&	47.2831(2)	&	547.258(2)	&				&	0.39(3)	&				&				&	0.73(1)	&			&			&	5.1	&		& f$_{\rm 7}$ in DM17, f$_{\rm 3}$ in CK03b	\\
F8	&	47.43924(5)	&	549.0653(6)	&	1.38(5)		&	1.45(3)	&	1.24(3)		&	0.786(6)	&	0.436(3)&	0.816(4)&	9.05	&	20.7&	24.2& f$_{\rm 1}$ in DM17, f$_{\rm 1}$ in CK03b	\\
F9	&	48.9185(3)	&	566.186(4)	&				&	0.22(3)	&				&				&	0.38(2)	&			&			&	5.3	&		& f$_{\rm 8}$ in DM17	\\
F10	&	49.4161(4)	&	571.946(5)	&				&	0.17(3)	&				&				&	0.43(3)	&			&			&	4.3	&		&			\\
F11	&	50.49210(8)	&	584.399(1)	&	0.55(5)		&	0.84(3)	&	1.06(3)		&	0.07(2)		&	0.022(5)&	0.145(4)&	5.33	&	19.9&	23.9	& f$_{\rm 2}$ in DM17		\\
F12	&	50.8312(4)	&	588.324(5)	&				&	0.15(3)	&				&				&	0.24(3)	&			&			&	3.9	&		& f$_{\rm 14}$ in DM17		\\
F13	&	53.6917(2)	&	621.431(2)	&	1.12(5)		&	0.40(3)	&	1.27(3)		&	0.797(8)	&	0.75(1)	&	0.534(4)&	10.43	&	9.5	&	16.38	&	f$_{\rm 3}$ in DM17		\\
F14	&	53.8915(6)	&	623.744(7)	&	0.64(5)		&			&				&	0.11(1)		&			&			&	6.55	&		&		& f$_{\rm 23}$ in DM17		\\
F15	&	54.2372(2)	&	627.745(2)	&	0.40(5)		&	0.44(3)	&	0.51(3)		&	0.15(2)		&	0.63(1)	&	0.002(9)&	4.55	&	10.3&	10.08	& f$_{\rm 4}$ in DM17		\\
\hline
\hline
\multicolumn{1}{c}{No.} & \multicolumn{2}{c}{Frequency}  & \multicolumn{1}{c}{ } & \multicolumn{1}{c}{$A_B$ 2016}    & \multicolumn{1}{c}{ } & \multicolumn{1}{c}{ } & \multicolumn{1}{c}{$\phi_B$ 2016}  & \multicolumn{1}{c}{ } & \multicolumn{1}{c}{ } & \multicolumn{1}{c}{S/N$_B$ 2016} & \multicolumn{1}{c}{ } & \multicolumn{1}{c}{cross ID} \\
\multicolumn{1}{c}{\#} &  \multicolumn{1}{c}{[\cd]}  &  \multicolumn{1}{c}{[$\mu$Hz]}   & \multicolumn{1}{c}{ } & \multicolumn{1}{c}{[mmag]}   & \multicolumn{1}{c}{ } & \multicolumn{1}{c}{ } & \multicolumn{1}{c}{ }   & \multicolumn{1}{c}{ }  & \multicolumn{1}{c}{ } & \multicolumn{1}{c}{ } & \multicolumn{1}{c}{ } & \multicolumn{1}{c}{ }  \\
\hline
F3	&	39.0629(1)	&	452.117(1)	&		&	0.75(11)	& 	&		&	0.47(2) 	&	&		&	4.4	 	&	& f$_{\rm 5}$ in DM17		\\
F4	&	43.5268(4)	&	503.783(4)	&		&	1.00(11) 	&	&		&	0.67(2)		&	&		&	4.7	  	&	& f$_{\rm 9}$ in DM17		\\
F8	&	47.43924(5)	&	549.0653(6)	&		&	2.30(11) 	&	&		&	0.749(8)	& 	&		&	13.4  	&	& f$_{\rm 1}$ in DM17, f$_{\rm 1}$ in CK03b	\\
F11	&	50.49210(8)	&	584.399(1)	&		&	0.95(11) 	&	&		&	0.69(2)		& 	&		&	5.8	  	&	& f$_{\rm 2}$ in DM17		\\
\hline
\end{tabular}
\end{scriptsize}
\end{center}
\tablefoot{The upper part of the table lists the pulsational properties derived from the BRITE red filter data (denoted with $R$), the lower part of the table those of the BRITE blue filter observations (denoted with $B$). Frequency, amplitude and phase errors are calculated using the relations by \citet{montgomery99}. References: DM17: \citet{mekarnia17}, CK03b: \citet{koen2003b}}
\end{table*}

\subsection{Analysis of the bRing photometry}

The complete bRing light curve used for the present analysis has a total time base of more than 620 days (panel e in Fig. \ref{all-lcs}). The Nyquist frequency lies at 135.37\,\cd. 
We could identify six of the previously reported pulsation frequencies (i.e., F2, F6, F8, F11, F13 and F15) from this data set (left side in Fig. \ref{bring-smei-amp}). The residual noise level after subtracting all formally significant frequencies is at 110\,ppm (Table \ref{tab:obs} and panel c in Fig. \ref{bring-smei-amp}). Panel a in Figure \ref{bring-smei-amp} shows the bRing spectral window.

\begin{table*}[htb]
\caption{Pulsation frequencies, amplitudes, phases and signal-to-noise values derived from the SMEI and bRing data sorted by increasing frequency and comparison to the literature.}
\label{tab:smei_bRing}
\begin{center}
\begin{tabular}{llllrrccl}
\hline
 \multicolumn{1}{c}{BRITE No.} & \multicolumn{1}{c}{Frequency} & \multicolumn{1}{c}{$A_{\rm SMEI}$} & \multicolumn{1}{c}{$A_{\rm bRing}$} & \multicolumn{1}{c}{$\phi_{\rm SMEI}$} & \multicolumn{1}{c}{$\phi_{\rm bRing}$} & \multicolumn{1}{c}{S/N$_{\rm SMEI}$}  & \multicolumn{1}{c}{S/N$_{\rm bRing}$}& \multicolumn{1}{c}{cross ID}   \\
 \multicolumn{1}{c}{[\cd]}  &  \multicolumn{1}{c}{[\cd]}   & \multicolumn{1}{c}{ }  & \multicolumn{1}{c}{[mmag]} & \multicolumn{1}{c}{ } & \multicolumn{1}{c}{ } & \multicolumn{1}{c}{ } & \multicolumn{1}{c}{ }    \\
\hline
	F1	&	34.39059(4)		&	0.36(8)	& --		&	0.0(2)	& --	&	4.1	& -- &	close to f$_{10}$ in DM17	\\
	F3	&	39.06307(3)		&	0.55(8)	& 0.8(2)		&	$-$0.3(1)	& 0.32(4)	&	5.5	& 5.1 &	f$_5$ in DM17	\\
	Fc	&	44.68351(2)		&	0.67(8)	& --	&	$-$0.1(1)	& --	&	6.8	& -- &	f$_{13}$ in DM17	\\
 F6     &  46.5428(4)       & -- & 0.5(2) & -- & 0.13(4) & -- & 5.8 & f$_{\rm 6}$ in DM17 \\
	F7	&	47.28348(4)		&	0.43(8)	 & --	&	$-$0.3(2)	& --	&	4.5	& -- &	f$_7$ in DM17	\\
	F8	&	47.43920(2)		&	0.76(8) & 1.1(2)		&	$-$0.4(1)	& 0.05(3)	&	7.8	& 6.9 &	f$_1$ in DM17	\\
	F11	&	50.49182(4)		&	0.36(8)	& 0.8(2)	&	0.4(2)	& 0.23(3)	&	3.9	& 5.3 &	f$_2$ in DM17	\\
	F13	&	53.67090(3)		&	0.45(8) & 1.1(2)		&	0.1(1)	& 0.69(3)	&	4.0	& 6.5 &	f$_3$ in DM17	\\
 F15 & 54.2372(3) & -- & 0.6(2) & -- & 0.94(4) & -- & 6.8 & f$_{\rm 4}$ in DM17 \\
\hline
\end{tabular}
\end{center}
\tablefoot{"BRITE No.'' lists the frequency numbers given in Table \ref{tab:freqs}. Frequency, amplitude and phase errors are calculated using the relations by \citet{montgomery99}. Reference in column cross-ID: DM17: \citet{mekarnia17}.}
\end{table*}

\subsection{Analysis of the SMEI photometry}

The SMEI light curves are affected by strong instrumental effects, such as large yearly flux fluctuations.
We corrected for this one-year periodicity of instrumental origin by phasing the raw data with a one-year period, calculating median values in 200 phase intervals, interpolating between these points and subtracting the interpolated light curve. In the next step we detrended the data repeatedly with simultaneous sigma clipping to remove outliers and suppress any instrumental signal at low frequencies. The detrending was done 30 times starting with a time interval to calculate the mean, T, of 100 days and a sigma of 5, and ending the procedure with T = 0.7 days and a sigma of 4. 
As a consequence of this method, frequencies lower than 0.5\,\cd\ are suppressed.
The annual light curves of \bpic\, show very similar behaviour, hence the corresponding minor differences were subtracted by detrending.
Panel f in Figure \ref{all-lcs} shows the complete reduced SMEI light curve.

The sampling of the SMEI data is only about one data point per 1.7 hours (i.e., the orbital period of the satellite) which results in a Nyquist frequency of only 7.08\,\cd. Hence, investigating SMEI data for the presence of \dsct-type pulsations in the range from about 30 to 70\,\cd\ as expected for \bpic\ goes already beyond the Nyquist frequency, $f_{\rm Nyquist}$. 
As it was shown, e.g., for Kepler data by \citet{murphy2013}, it is similarly possible to do super-Nyquist asteroseismology using the SMEI data as real peaks remain as singlets even if they are above $f_{\rm Nyquist}$. Panel e in Figure \ref{bring-smei-amp} shows the complete amplitude spectrum of the SMEI data ranging up to 100\,\cd. It can clearly be seen how the pulsation frequencies between 30 and 60\,\cd remain single peaks that can be easily distinguished from aliases caused by $f_{\rm Nyquist}$. The dips in the noise around 42.5 and 52\,\cd\ are caused by the strong detrending.

The frequency analysis of the SMEI data yielded seven pulsation frequencies that are either present in the BRITE data or were reported by \citet{mekarnia17} or both (Table \ref{tab:smei_bRing}, panels d to f in Figure \ref{bring-smei-amp}).

\section{Spectropolarimetric analysis}

\subsection{Stokes profiles}

\begin{figure}
\resizebox{\hsize}{!}{\includegraphics[clip]{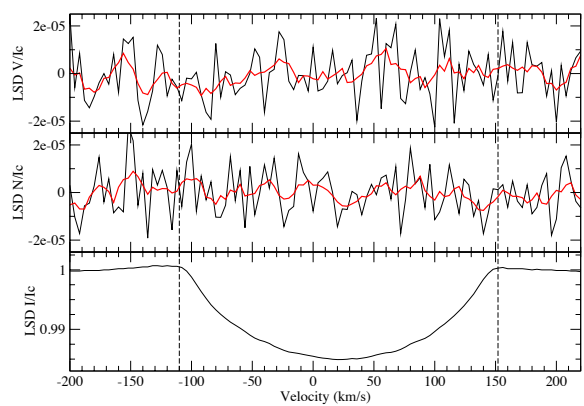}}
\caption[]{LSD Stokes V (top), N (middle) and Stokes I (bottom) profiles of
$\beta$\,Pic. Data are shown in black while red lines show a smoothed profile.
Vertical dashed lines indicate the range of integration to calculate the
longitudinal magnetic field and estimate the FAP.}
\label{betpic_lsd}
\end{figure}

We applied the Least Squares Deconvolution (LSD) method \citep{donati1997} to
the HARPSpol data to produce seven mean LSD Stokes I, Stokes V, and N profiles. The
seven sequences were then co-added to produce one final set of LSD profiles, shown
in Fig.~\ref{betpic_lsd}. 

Performing LSD requires the use of a mask indicating the list of lines present
in the spectrum and to be used in the averaging, their wavelengths, depths and
Land\'e factors. To produce this mask, we started from a line list extracted from
the {\sc VALD3} atomic database \citep{piskunov1995, kupka1999} for a star with
$T_{\rm eff}$ = 8200 K and $\log g$ = 4.0 (i.e., the stellar parameters of $\beta$ Pic taken from \citep{lanz1995}. We retained only the lines with a predicted
depth larger than 0.01. In addition, we rejected hydrogen
lines, lines blended with H lines or interstellar lines, and regions affected by
telluric absorption. Finally, we adapted the depth of the lines in the mask to
the actual depth of the lines observed in the spectra with an automatic fitting
routine. In total, we used 6391 spectral lines, with a mean wavelength of
503.8 nm and a mean Land\'e factor of 1.202. 

The LSD N profile represents the Null polarisation and is used as a sanity check
for the spectropolarimetric measurement. The fact that the N profile shows only
noise (see Fig.~\ref{betpic_lsd}, middle panel) indicates that the
spectropolarimetric measurement has not been polluted by instrumental effects or
stellar variability of non-magnetic origin.

The LSD Stokes V profile should indicate a Zeeman signature if a magnetic field
was present in $\beta$\,Pic. Since this profile also shows only noise (see
Fig.~\ref{betpic_lsd}, top panel), it indicates that $\beta$\,Pic is not
magnetic, at the level of precision of our measurement.  

We can evaluate the detection of a magnetic field statistically with the False
Alarm Probability (FAP). We check the presence of a signature in the LSD Stokes
V profile inside the velocity range [$-$110:152] km~s$^{-1}$, compared to the mean
noise level in the LSD Stokes V profile outside the line. We adopted the
convention defined by \cite{donati1997} that there is a definite or marginal
magnetic detection if FAP $<$ 0.1\%.

The FAP analysis of the LSD Stokes V profile leads to no magnetic detection
(with a FAP = 99.99\%). A similar analysis of the N profile also indicates no
detection in N (as expected).

\subsection{Longitudinal magnetic field measurement}

A first quantitative estimate of the non-detection of a magnetic field in
$\beta$\,Pic can be obtained via the measurement of the (undetected)
longitudinal magnetic field $B_l$. 

To this aim, we used a center-of-gravity method \citep{rees1979,wade2000} and
applied it to the Stokes V and N profiles. We integrated the profiles over the
same velocity range as for the FAP analysis, i.e. [$-$110:152] km~s$^{-1}$. We
obtained $B_l$ = $-$14 $\pm$ 20 G. We applied the same method to the N profile and
obtained $N_l$ = 11 $\pm$ 20 G. Both values are compatible with 0, indicating
that no field is detected in $\beta$\,Pic. The error of 20 G shows that our
measurement has a good precision.

\subsection{Upper limit of the undetected magnetic field}

Another quantitative estimate of the non-detection of a magnetic field in
$\beta$\,Pic can be obtained by determining the maximum strength of a magnetic
field that could have remained hidden in the noise of our data. 

Since $\beta$\,Pic is an A6V star, its envelope is radiative and thus, if it
hosts a magnetic field, it must be of fossil origin \citep{neiner2015}. Fossil
fields are usually dipolar and tilted with respect to the rotation axis of the
star \citep{grunhut2015}. Therefore, to estimate the upper limit of a possibly
undetected magnetic field, we assumed an oblique dipolar field.

We followed the method described in \cite{neiner_pacwb2015}: for various values
of the polar magnetic field strength $B_{\rm  pol}$, we calculated 1000 models
of the LSD Stokes $V$ profile with random inclination angle $i$, obliquity angle
$\beta$, and rotational phase, and a white Gaussian noise with a null average
and a variance corresponding to the SNR of the observed profile. We first fitted
the LSD $I$ profile, we then calculated local Stokes $V$ profiles assuming the
weak-field case and we integrated over the visible hemisphere of the star. We
used a projected rotational velocity of $v \sin i$ = 124 km~s$^{-1}$ taken from \citep{lanz1995} and a limb-darkening coefficient of 0.6. In
this way we obtained a synthetic Stokes $V$ profile for each model, which we
normalised to the intensity continuum. We used the same Land\'e factors and
wavelengths as in the LSD calculation.

\begin{figure}
\resizebox{\hsize}{!}{\includegraphics[clip]{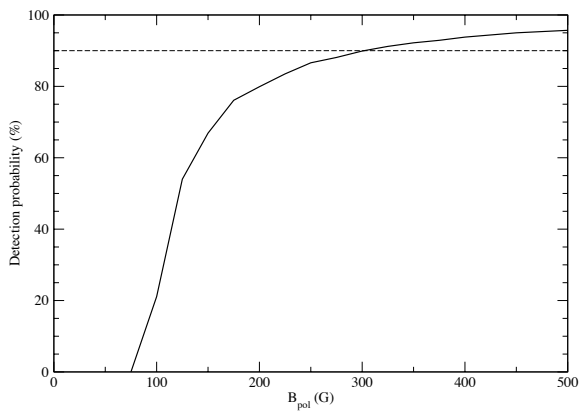}}
\caption[]{Detection probability of a magnetic field in the observation of
$\beta$\,Pic as a function of the magnetic polar field strength. The horizontal
dashed line indicates the 90\% detection probability. A dipole field stronger than
$B_{\rm  pol}$ = 300 G would have statistically been detected.}
\label{betpic_limit}
\end{figure}

We then computed the probability of detection of a field in these 1000 models by
applying the Neyman-Pearson likelihood ratio test. We further calculated the
rate of detections among the 1000 models depending on the field strength (see
Fig.~\ref{betpic_limit}). We required a 90\% detection rate to consider that the
field should have statistically been detected. This translates into an upper
limit for the possible undetected dipolar field strength for $\beta$\,Pic of
$B_{\rm  pol}$ = 300 G. Using a 50\% detection rate would bring the limit of the
possible undetected dipolar field strength down to $B_{\rm  pol}$ = 120 G.

\section{Amplitude variability}

\dsct stars can show variable pulsation amplitudes that are either of intrinsic (beating of unresolved frequencies, non-linearity or mode-coupling) or extrinsic (binarity and multiple systems) origin \citep[for a detailed overview see][]{bowman2016}. We examined the amplitude variability of \bpic\ using the three seasons of BRITE-Constellation red filter data and the bRing observations.

\subsection{Annual changes in the amplitudes}
\label{sect.ampvar}
\begin{figure}
\begin{center}
\includegraphics[width=0.49\textwidth]{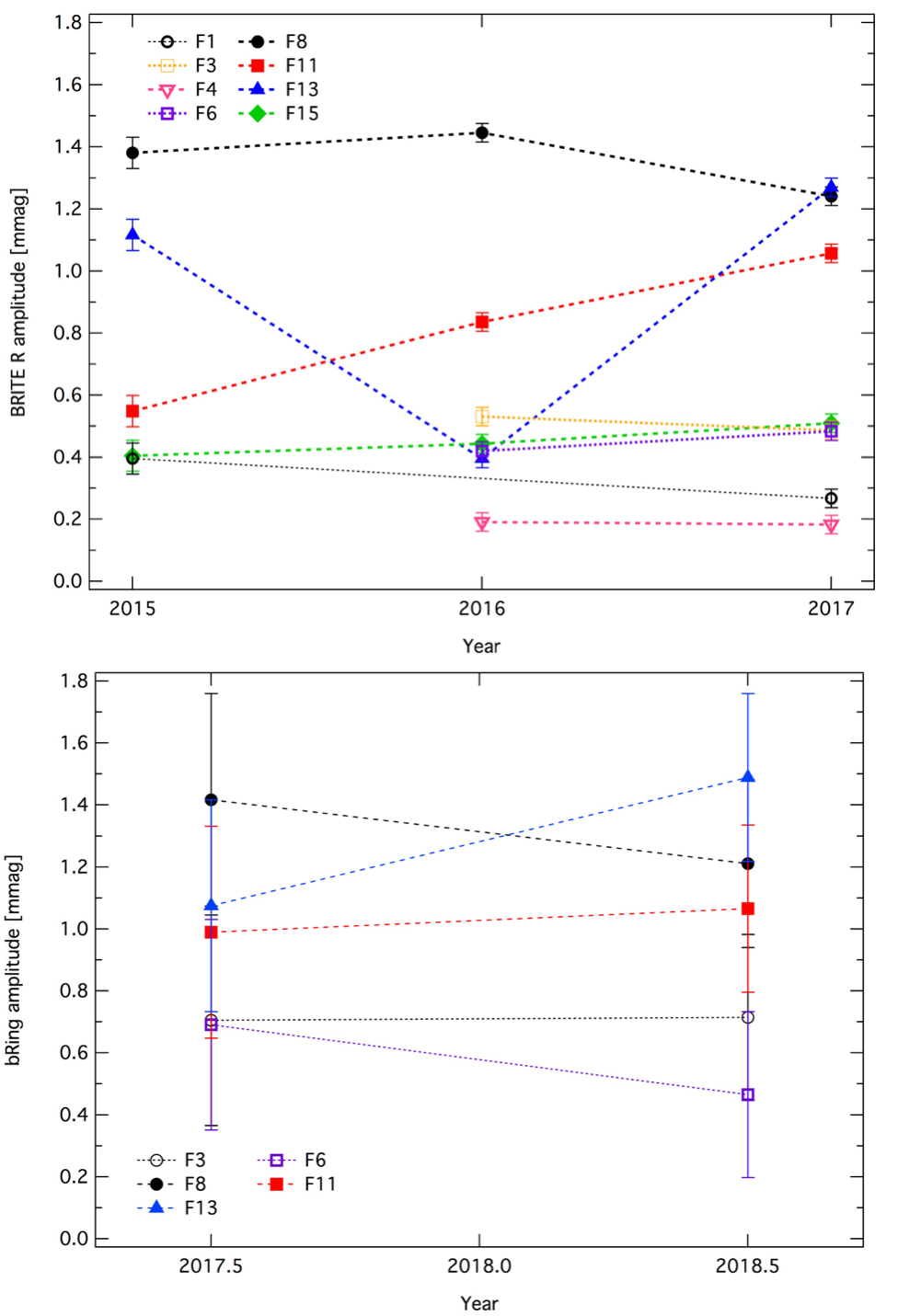}
\caption{Top panel: Annual behavior of the amplitudes of the four pulsation frequencies F8, F11, F13 and F15 present in all three BRITE red filter data sets (filled symbols) and of the four pulsation frequencies F1, F3, F4 and F6 that appear in two of the three BRITE R epochs (open symbols). Bottom panel: Annual behaviour of the five pulsation frequencies F3, F6, F8, F11, and F13 in the bRing data.}
\label{ampvar_annual}
\end{center}
\end{figure}

\begin{figure}
\begin{center}
\includegraphics[width=0.49\textwidth]{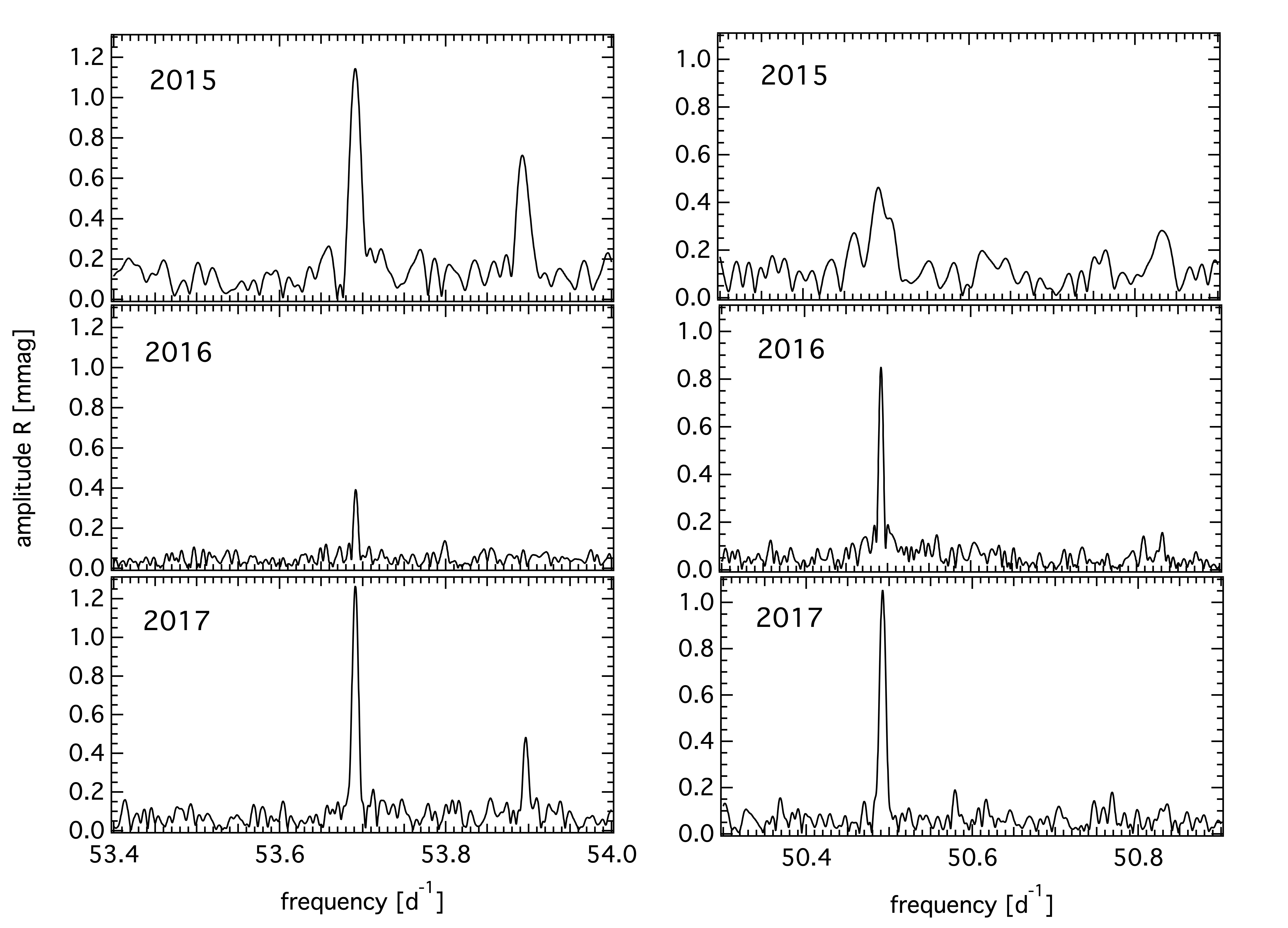}
\caption{Zoom around F13 (left side, seen alongside with F14 in the 2015 data to the right of F13) and F11 (right side) into the original amplitude spectra of the BRITE red filter data obtained in 2015 (top), 2016/2017 (middle) and 2017/2018 (bottom). The peak next to F13 in the bottom left panel showing the 2017/18 data is an alias frequency to F3 with the orbital frequency of BHr (i.e., 14.83053\cd) that only appears quite close to the location of F14 which itself is not found in the 2017/18 data.}
\label{F11-F13}
\end{center}
\end{figure}

Using the three BRITE R filter data sets obtained in consecutive years and the bRing light curve, we can study the annual amplitude variability of \bpic's pulsation frequencies from 2015 to 2018. Only four of the 15 identified pulsation frequencies appear in the BRITE data sets of all three years: F8, F11, F13 and F15. Frequencies F1, F3, F4, and F6 are detectable in two of the three BRITE observing seasons and the other seven frequencies only appear in one year of BRITE data. The top panel in Figure \ref{ampvar_annual} illustrates that the amplitudes of six of these frequencies remain rather stable or change only slightly from year to year (i.e., F1, F3, F4, F6, F8 and F15), while F11 shows a strong increase in amplitude and F13's amplitude decreases significantly in the 2016/2017 observations and increases again in the 2017/2018 data set. 
A zoom into the BRITE R filter amplitude spectra around F11 and F13 illustrates this behaviour in the Fourier domain within the three seasons of BRITE-Constellation observations (Figure \ref{F11-F13}).

In a next step, we divided the 620-day long bRing light curve into two parts of equal length and studied the resulting behaviour of the amplitudes with respect to the center points in time, i.e., mid 2017 and mid 2018. Despite the higher noise level in the bRing data which translates into larger errors on the amplitudes, it is obvious that the amplitude for F13 increases during this period of time. As the errors of the other four pulsation amplitudes (i.e., F3, F6, F8, and F11) are quite large, no clear conclusion on variability or stability can be drawn from these data.

\subsection{Amplitude variability within observing seasons}

\citet{mekarnia17} investigated the presence of amplitude and phase changes for their 10 first frequencies during their 7-month long observations by dividing their data set into seven parts, each 30 days in length. They showed that only their frequency f$_3$ (corresponding to our frequency F13) at 53.69138\cd\, changes its amplitude from 403 to 826$\pm$66 ppm (i.e., 0.403 to 0.826$\pm$0.066 mmag).

\begin{figure}
\begin{center}
\includegraphics[width=0.49\textwidth]{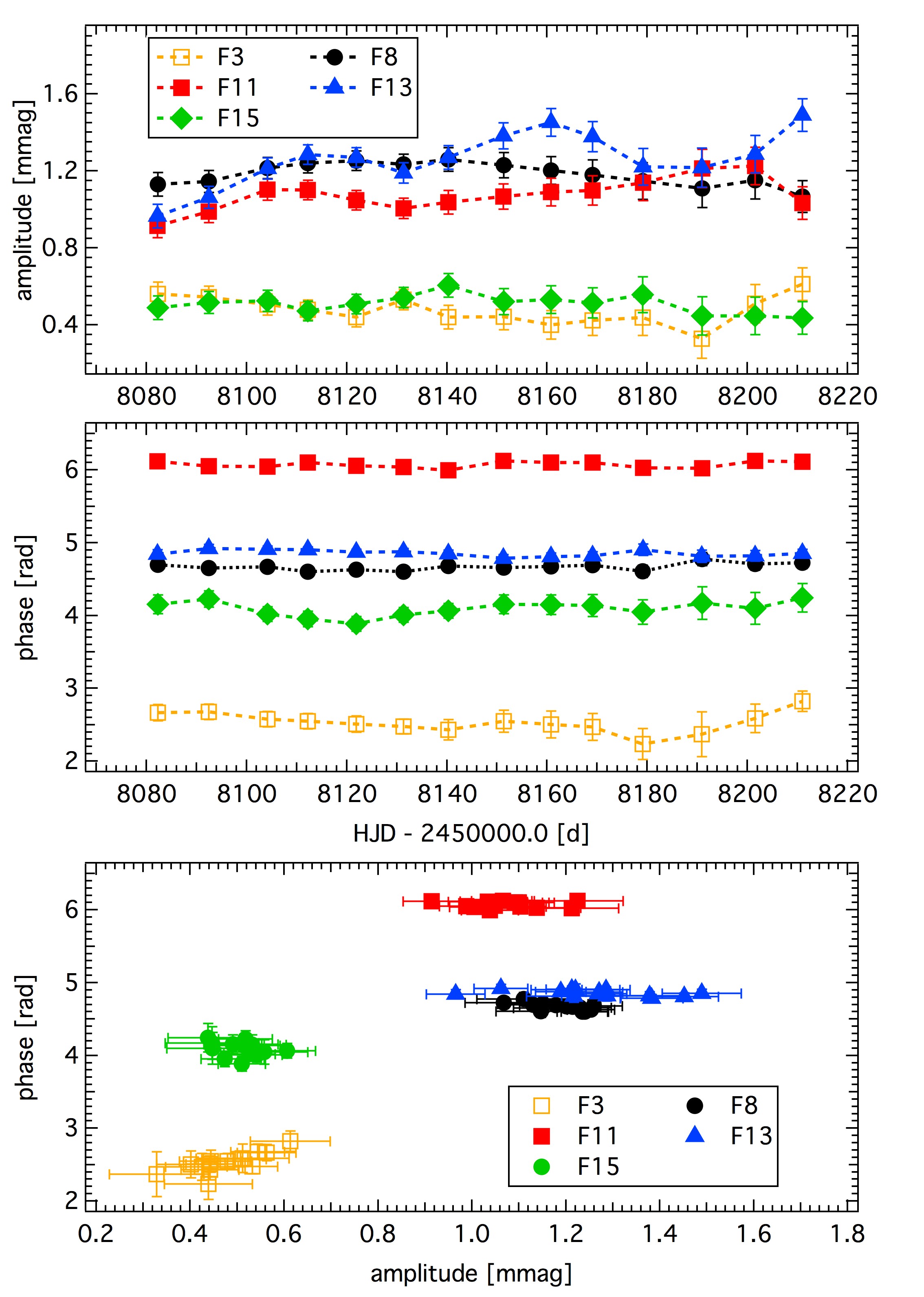}
\caption{Amplitude (top panel) and phase (middle panel) behaviour of the five highest-amplitude pulsation frequencies during the 2017/18 BHr observations calculated from 30-day subsets with 20-day overlaps, and amplitude-phase relation (bottom panel). The errors in phases are often smaller than the symbol size.}
\label{ampvar_5freqs}
\end{center}
\end{figure}

For the purposes of comparison, we conducted the same analysis as \citet{mekarnia17}, i.e., calculating the amplitude behaviour using 30-day subsets. As the overall amplitude of our frequency F13 is significantly lower during the longest BRITE observing run in 2016/17 (see Figure \ref{ampvar_annual}) and, therefore, is buried in the noise (i.e., not significant) when calculating 30-day subsets, we chose the 2017/18 BHr data set for this comparative analysis. Using subsets of 30-days length with 20 days of overlap, we find the amplitude of F13 to increase from 965 ppm (i.e., 0.965 mmag) to a maximum of 1489 ppm (i.e., 1.489 mmag); see blue symbols and line in the top panel of Figure \ref{ampvar_5freqs}).

As the ASTEP observations were conducted in the time from March to September 2017 \citep{mekarnia17}, there is an overlap of about four months with the second BRITE data set from 2016/17 which was taken a bit earlier starting in November 2016 and running until June 2017. The overall amplitude of our F13 at 53.6917\,\cd\ in this data set is only 396 ppm (i.e., 0.396 mmag), being the lowest in our analysis. During the subsequent observations with ASTEP the amplitude seems to have already increased. When BRITE-Constellation picked up \bpic\ again in November 2017 for the third season, the amplitude continued to increase. This effect is evident despite the fact that ASTEP and BHr are different instruments carrying different filters. 

The other four highest-amplitude frequencies -- F3, F8, F11 and F15 -- vary to a much smaller extent or remain basically at a constant level during this season (top panel in Figure \ref{ampvar_5freqs}). These four frequencies have been also identified in \citet{mekarnia17}, but not marked as showing variable amplitudes.

The pulsation phases for the five selected frequencies during the BRITE 2017/18 observations can clearly be regarded as stable (middle panel in Figure \ref{ampvar_5freqs}). As the changes in the amplitudes of \bpic\, are not correlated with changes in phases (bottom panel in Figure \ref{ampvar_5freqs}), we interpret the variability of the amplitudes as being intrinsic and not caused by beating of two or more unresolved modes.

In a final test we used the bRing data set for a comparable investigation of amplitude variability. Due to the higher noise in the data compared to the BRITE observations, we had to choose 100-day subsets with 50-day overlaps to detect frequencies F8, F11, and F13. Unfortunately, the uncertainties of the amplitudes derived in the 100-day subsets are too high for an analogous interpretation of amplitude variability which can be seen in Figure \ref{ampvar_bring}. 

\section{Asteroseismic interpretation}

\begin{figure}[htb]
\begin{center}
\includegraphics[width=0.5\textwidth]{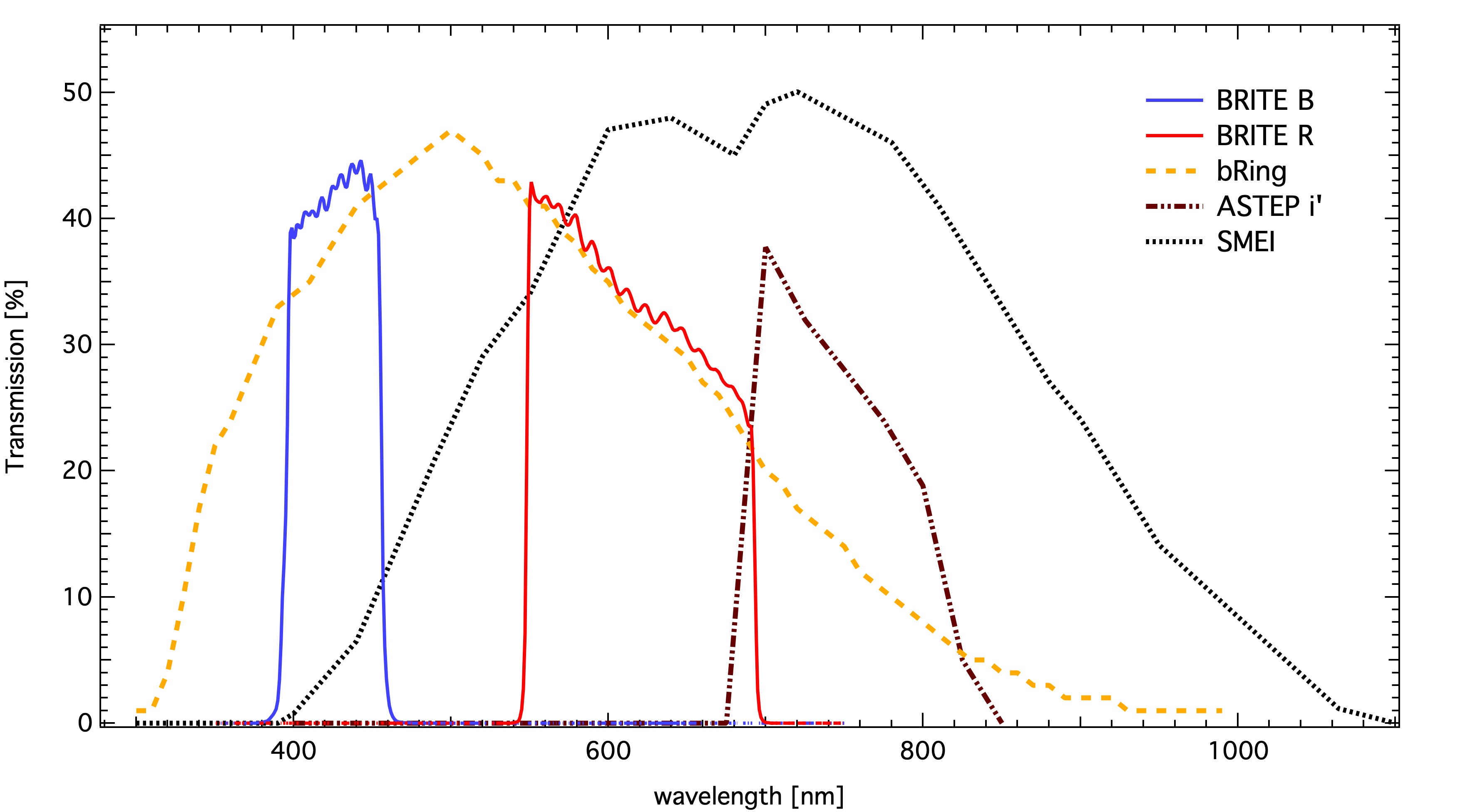}
\caption{Transmission curves of the five instruments: BRITE B (blue solid line), BRITE R (red solid line), bRing (orange dashed line), ASTEP $i'$ (dark red dashed-dotted line) and SMEI (black dotted line).}
\label{transmissions}
\end{center}
\end{figure}

We used the pulsation frequencies and amplitudes derived from up to five passbands to identify the pulsation modes from comparing the observed and theoretical  normalised amplitudes. The five filters are BRITE B, BRITE R, SMEI, and bRing (filter information is given in Table \ref{tab:obs}) together with the previously published 31 frequencies using the ASTEP instrument \citep{mekarnia17} which uses a Sloan $i^{\prime}$ filter (passband from 695 to 844 nm). The transmission curves of the filters of all five instruments  used in the present analysis are illustrated in Figure \ref{transmissions}.

In order to carry out mode identification we used a sequence of seven $1.8\,M_{\odot}$ models from the Self-Consistent Field method \citep{jackson2005, macgregor2007} with rotation rates ranging from $0$ to $0.6\,\OmegaK$ in increments of $0.1\,\OmegaK$, where $\OmegaK=\sqrt{GM/R^3}$ is the Keplerian break-up rotation rate.  
The initial mass for the models was chosen based on the value of $1.8\,M_{\odot}$ given by \citet{wang2016}.
The models are Zero Age Main Sequence (ZAMS) models which is fairly realistic in the case of $\beta$ Pic which was identified as a 23 Myr-old ZAMS star \citep{mamajek2014}.  We then calculated their low-degree acoustic mode pulsations for $n$ ranging from $1$ or $2$ to $10$, $\ell$ from $0$ to $3$, and $m$ from $-3$ to $3$, using the adiabatic version of TOP \citep[Two-dimension Oscillation Program, see][]{reese2006,reese2009}.  Pseudo non-adiabatic mode visibilities (i.e., disk-integrated geometric factors) were derived using the approach described by \citet{reese2013,reese2017} where the non-adiabatic pulsation amplitudes and phases came from 1D pulsation calculations using the MAD code \citep{dupret2001} in non-rotating stellar models which span the relevant range in effective temperature and gravity.  These mode visibilities were calculated for all inclinations between $0^{\circ}$ and $90^{\circ}$ in increments of $1^{\circ}$.  The intensities in different photometric bands at each point of the stellar surface were obtained by integrating a black body spectrum at the latitude-dependent effective temperature multiplied by the filter's transmission curve, thus taking gravity darkening into account.  Limb darkening was included by multiplying these intensities by a Claret law \citep{claret2000} for the filter\footnote{The relevant transmission curves were downloaded from: \url{http://www.aip.de/en/research/facilities/stella/instruments/data/johnson-ubvri-filter-curves}} which was the closest match to the photometric bands used here, as listed in Table~\ref{tab:Claret_filters}.  The correlations given in the third column of Table~\ref{tab:Claret_filters} are defined as $\int_0^{\infty} f_1(\lambda) f_2(\lambda) \mathrm{d}\lambda/\sqrt{\int_0^{\infty} f_1^2(\lambda)  \mathrm{d}\lambda \int_0^{\infty} f_2^2(\lambda) \mathrm{d}\lambda}$, where $\lambda$ is the wavelength.

\begin{table}[htbp]
\begin{center}
\begin{tabular}{ccc}
\hline
This study &
Claret (2000) &
Correlation \\
\hline
BRITE red  & Johnson R & 0.887 \\
BRITE blue & Johnson B & 0.857 \\
ASTEP $i^{\prime}$ & Johnson I & 0.445 \\
SMEI       & Johnson R & 0.729 \\
bRing      & Johnson V & 0.609 \\
\hline
\end{tabular}
\end{center}
\caption{Correspondence between photometric bands used to observe $\beta$ Pic (``This study'')
and those used to implement the limb-darkening law (``Claret (2000)'').  The third column gives
the degree of correlation between the two. \label{tab:Claret_filters}}
\end{table}

As a first step, we searched for the low-degree modes that \textit{individually} provide the best match to the observed amplitudes.  We did not attempt to match observed phase differences as the pseudo non-adiabatic calculations were not considered to be sufficiently reliable to provide accurate theoretical phase differences.  We chose not to compare normalised amplitudes directly because this amounts to choosing one of the photometric bands as a reference band and normalising the amplitudes in the other bands with respect to the amplitude in this reference band.  This can lead to difficulties if the amplitude in the chosen reference band is close to zero.  Instead, we normalised the observed amplitudes so that the sum of their squares equals one.  For the sake of consistency, the errors on the amplitudes are also normalised by the same factor.  The theoretical amplitudes are then normalised so as to optimize the $\chi^2$ fit to the observations taking into account the errors.  Also, given the relatively large increment on the rotation rate, the normalised amplitudes and frequencies were interpolated to intermediate rotation rates in increments of $0.01\,\OmegaK$.

\begin{figure*}[htb]
\includegraphics[width=\textwidth]{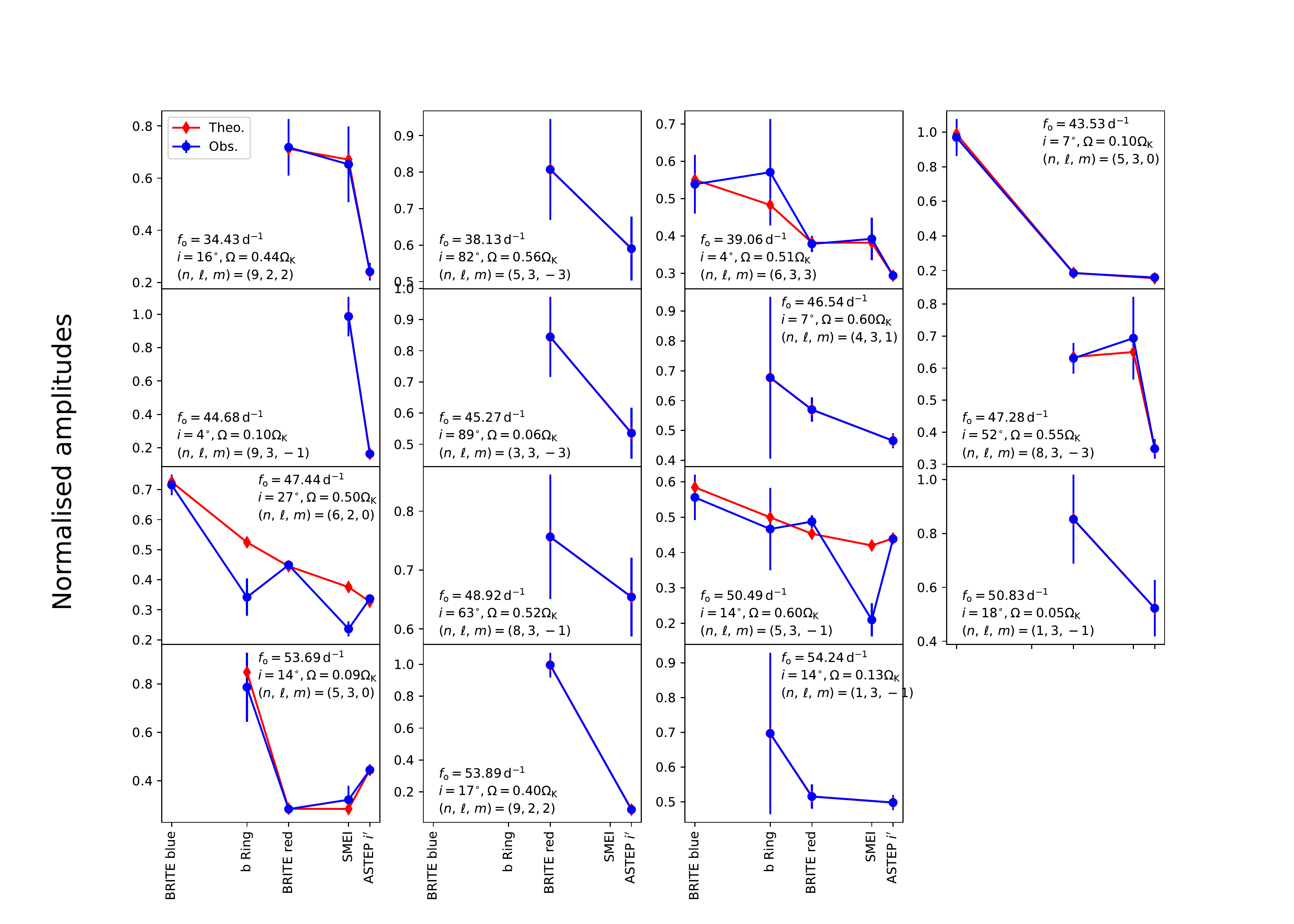}
\caption{Observed (blue) and best-fitting theoretical (red) normalised amplitudes using individual fits. $f_o$  represents the observed frequencies.  In a number of cases, the match between observed and theoretical normalised amplitudes is sufficiently good that the latter is hidden in the plots.  Finally, the total $\chi^2$ value is 68.9.}
\label{fig:best_individual}
\end{figure*}

\begin{figure}[htb]
\includegraphics[width=0.5\textwidth]{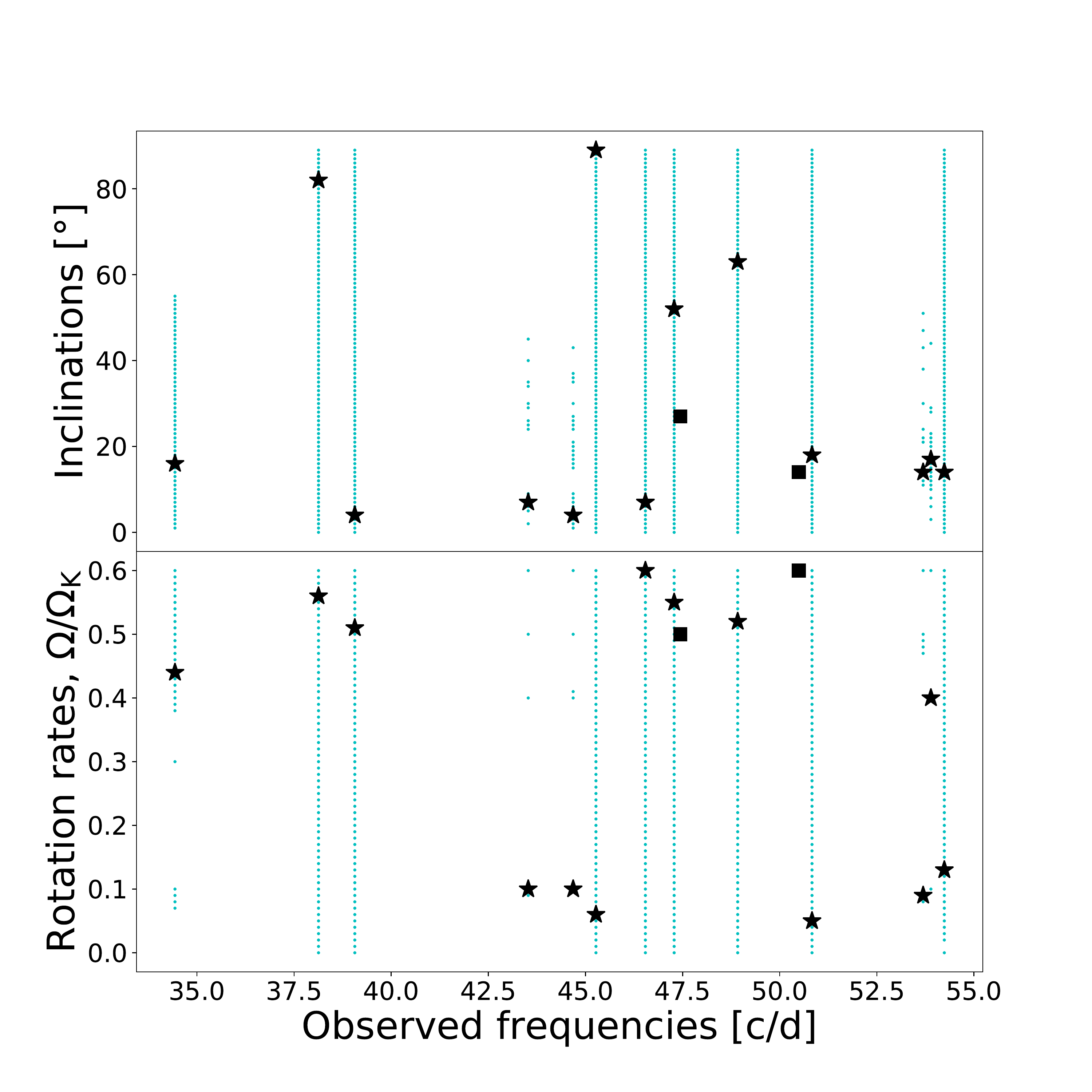}
\caption{Inclinations (upper panel) and rotation rates (lower panel) obtained for each mode fit individually. The dark blue symbols represent the best solutions (i.e., those illustrated in Figure \ref{fig:best_individual}).  The small light blue dots are other solutions that satisfy the criterion $\chi^2 \leq (N_{bands} - 1)$, where $\chi^2$ is the $\chi^2$-value on the amplitudes for that particular mode and $N_{bands}$ is the number of bands in which that mode is detected. Hence, the light blue solutions give an idea of the uncertainties on the above results.}
\label{fig:parameters_individual}
\end{figure}

Figure~\ref{fig:best_individual} shows a comparison between the observed and best-fitting normalised amplitudes.  A good match can be observed for most of the modes and a relatively low $\chi^2$ value is obtained.  However, since the modes were fit individually, the inclinations and rotation rates obtained for the different modes do not match.  
Fig.~\ref{fig:parameters_individual} illustrates these results: dark blue symbols represent the best solutions (i.e., those illustrated in Fig.~\ref{fig:best_individual}).  The small light blue dots are other solutions that satisfy the criterion $\chi^2 \leq (N_{bands} - 1)$, where $\chi^2$ is the $\chi^2$-value on the amplitudes for that particular mode and $N_{bands}$ is the number of bands in which that mode is detected. On the right side of the latter relation, $(N_{bands} - 1)$ was chosen and not $N_{bands}$ as the intrinsic amplitude of theoretical mode is a free parameter. If none of the solutions (including the best solution) satisfies the above criterion, then the best solution is plotted using a square rather than a star. Hence, the light blue solutions give an idea of the uncertainties on these solutions.
Hence, in order to obtain a more coherent result it is necessary to fit the modes \textit{simultaneously} using a fixed value of the inclination and rotation rate.  The fit to the normalised amplitudes shown here are nonetheless useful as they represent the best fit one can hope to achieve using our set of theoretical mode visibilities.

We then fit the frequencies and normalised amplitudes simultaneously.  In order to achieve this, we carried out Monte Carlo Markov Chains (MCMC) runs using the python \texttt{emcee} package \citep{foreman-mackey2013} and using the rotation rate, $\Omega/\OmegaK$, the inclination, $i$, and a dimensionless scale factor, $f$, on the frequencies as free parameters together with the free amplitudes of the 15 pulsation frequencies. These 18 free parameters are then used to fit 46 amplitudes in five passbands, 15 pulsation frequencies and four classic constraints (mass, radius, \logg, and \vsini) corresponding to in total 65 observables. We assumed a uniform prior on $\Omega/\OmegaK$ over the interval $[0.0,0.6]$ and on $f$ over the interval $[0.5, 2]$.  The prior on $i$ was proportional to $\sin i$ in accordance with what is expected for random orientations of the rotation axis, although as will be explained in what follows we sometimes restricted the range of $i$ values.  The scale factor was introduced to compensate for the fact that the set of stellar models was limited to a single mass.  Indeed, multiplying the frequencies by a dimensionless scale factor $f$ amounts to carrying out a homologous transformation in which the mean density of the model is multiplied by $f^2$ thus providing a poor substitute for modifying the mass.  This leads to the following relations:
\begin{equation}
f^2 = \frac{M}{R^3}\frac{R^3_{\mathrm{mod}}}{M_{\mathrm{mod}}}, \qquad
\OmegaK = f \OmegaKmod,
\end{equation}
where quantities with the subscript ``$\mathrm{mod}$'' refer to the model prior to scaling, and those without this subscript to the scaled model (which should hopefully be close to the actual properties of $\beta$ Pic).  We note that since our models only depend on $\Omega/\OmegaK$, this parameter also uniquely determines $\OmegaKmod$.  Finally, when matching the theoretical modes to the observed ones, we were careful to avoid assigning the same identification to two different observed modes.

Once the values of $\Omega/\OmegaK$, $f$, and $i$ are fixed, it is possible to deduce the radius of the scaled model using the observational constraints on $\logg$ and $\vsini$ (recalled in Table~\ref{tab:classic_constraints} below) using two different methods.  Indeed, one has the following relations:
\begin{eqnarray}
R &=& \frac{g}{\OmegaK^2}, \\
R &=& \frac{\vsini}{\OmegaK (\Omega/\OmegaK) \sin i},
\end{eqnarray}
where we recall that $\OmegaK = f \OmegaKmod$.  We note that we have neglected possible differences between the polar and equatorial radius in the above expressions.  Given that the values of $\Omega/\OmegaK$, $f$, and $i$ are selected by the MCMC algorithm, there is no reason to assume that the two above expressions give a priori the same values of $R$.  One could therefore eliminate one of the free parameters in the MCMC calculations.  We, however, opted for a different approach which consists in keeping all of the free parameters, calculating $R$ using both of the above equations, and rejecting solutions for which the two values of $R$ differ by more than $20\,\%$.  The final value of $R$ is then obtained by minimizing the following least-squares cost function:
\begin{equation}
J_1(R) = \left(\frac{R\OmegaK - g}{\sigma_g}\right)^2
     + \left(\frac{R\Omega\sin i - v \sin i}{\sigma_v}\right)^2
\label{eq:cost_R}
\end{equation}
The mass is subsequently obtained via the Keplerian breakup rotation rate and the above radius:
\begin{equation}
M = \frac{R^3\OmegaK^2}{G}
\end{equation}

At this point, it is useful to discuss the classical constraints on \bpic's fundamental atmospheric parameters.  In Table~\ref{tab:classic_constraints} we list various constraints found in the literature.  We note that other values have been obtained for some of these parameters: $\logg=4.15$ dex according to \citet{gray2006}, the angular diameter $\theta = 0.84 \pm 0.12 \pm 0.10$ mas including limb darkening according to \citet{difolco2004}, $\theta = 0.712 \pm 0.010$ mas based on surface-brightness relations from \citet{kervella2004} \citep[see][]{defrere2012} and $\vsini=130\pm 4$ \kms\ according to \citet{royer2002}.    It is then interesting to investigate to what extent these values are self-consistent.  For this, we minimized the following cost function in order to obtain a coherent set of parameters:
\begin{eqnarray}
J_2(\theta,\pi,M,\Teff) &=&
             \left(\frac{\theta-\theta_{\mathrm{obs}}}{\sigma_\theta}\right)^2
           + \left(\frac{\pi-\pi_{\mathrm{obs}}}{\sigma_\pi}\right)^2
           +  \left(\frac{\Teff-(\Teff)_{\mathrm{obs}}}{\sigma_T}\right)^2 \nonumber \\
           &+& \left(\frac{M-M_{\mathrm{obs}}}{\sigma_M}\right)^2
           + \left(\frac{\log_{10} \left( \frac{(2\pi)^2 GM}{\theta^2 d^2}\right)-\logg_{\mathrm{obs}}}{\sigma_{\logg}}\right)^2 \nonumber \\
           &+&  \left(\frac{4\pi\sigma \left(\frac{\theta d}{2\pi}\right)^2\Teff^4-L_{\mathrm{obs}}}{\sigma_L}\right)^2
\end{eqnarray}
where $d$ is one astronomical unit which is used to convert parallax into distance, quantities with the subscript ``$\mathrm{obs}$'' are the observed values from Table~\ref{tab:classic_constraints}, $\sigma_x$, $x$ representing any one of these quantities and the associated uncertainties, and $\sigma_M$ the average of the two errors on mass from the table.  As can be seen, the above cost function makes use of the relations:
\begin{equation}
L = 4 \pi \sigma R^2 \Teff^4, \qquad g = \frac{GM}{R^2}, \qquad R = \frac{\theta d}{2 \pi},
\end{equation}
where $\pi$ represents the parallax in the third equation.  The best fit is provided in the fourth column of Table~\ref{tab:classic_constraints}.  As can be seen, the observational constraints are mostly consistent.  This would have been less true if some of the alternate values such as $\logg=4.15$ dex \citep{gray2006} or $\theta=0.84\pm 0.12 \pm 0.10$ \citep{difolco2004} had been used.  The errors were estimated by carrying out 10000 Monte Carlo realizations of the observed errors before carrying out the above minimization and taking the standard deviation of the resultant parameters.  In some cases, these are very similar to the observed errors (e.g. the mass), whereas in other cases, they are considerably smaller (e.g. $\logg$).

\begin{table*}[htbp]
\begin{center}
\caption{List of classical constraints and least-squares fit. \label{tab:classic_constraints}}
\begin{tabular}{ccccc}
\hline
Quantity &
Value &
Reference &
Fit &
MCMC runs \\
\hline
Mass (\Msun) & $1.80^{+0.03}_{-0.04}$ & \citet{wang2016} & $1.797 \pm 0.035$ & Truncated Gaussian$^a$ $1.797 \pm 0.053$  \\ 
Parallax, $\pi$ (mas) & $51.44 \pm 0.12$ & \citet{vanleeuwen2007} & $51.45 \pm 0.12$ & -- \\ 
Angular diameter, $\theta$ (mas) & $0.736 \pm 0.019^{b}$ & \citet{defrere2012} & $0.716 \pm 0.012$ & -- \\ 
Radius (\Rsun) & $1.538 \pm 0.040$ & Deduced from $\pi$ and $\theta$ & $1.497 \pm 0.025$ & Truncated Gaussian$^a$ $1.538 \pm 0.040$  \\ 
$\logg$ (dex) & $4.25 \pm 0.10$ & \citet{lanz1995} & $4.343 \pm 0.017$ & Gaussian $4.25 \pm 0.10$ \\ 
Luminosity (\Lsun) & $8.47 \pm 0.23^{c}$ & \citet{crifo1997} & $8.62 \pm 0.21$ & -- \\ 
$\Teff$ (K) & $8143 \pm 67$ & Average from multiple papers$^d$ & $8090 \pm 59$ & -- \\ 
$\vsini$ (\kms) & $124 \pm 3$ & \citet{koen2003b} & --  & Gaussian $124 \pm 3$\\ 
\hline
\end{tabular}
\end{center}
\tablefoot{$^a$The truncated Gaussian distributions are truncated at $\pm 3\sigma$, i.e. solutions are rejected beyond this limit. $^b$ $\theta$ is the limb-darkened angular diameter \citep{defrere2012}. This error is obtained as the quadratic sum of the statistical and systematic errors given in \citet{defrere2012}.  $^c$ This is deduced from the value of $M_{\mathrm{bol}}$ rather than $L$ provided in \citet{crifo1997}, and assuming it has the same uncertainty as $M_v$. $^d$  The following values and errors were used in this average: 
  $7995$ K \citep[][using the calibration of \citealt{castelli1997}]{saffe2008},
  $8035 \pm 74$ K \citep{zorec2012},
  $8045 \pm 97$ K \citep{blackwell1998},
  $8052$ K \citep{gray2006},
  $8084$ K \citep{schroeder2009},
  $8128$ K \citep{AllendePrieto1999},
  $8157$ K \citep[][using the calibration of \citealt{napiwotzki1993}]{saffe2008},
  $8200$ K \citep{holweger1995},
  $8230 \pm 350$ K \citep{sokolov1995},
  $8300 \pm 282$ K \citep{david2015},
  $8500$ K \citep{mittal2015},
  $8543$ K \citep{dasilva2009}.
}
\end{table*}

In the MCMC runs that follow,  we used the observational constraints directly rather than using the results from the fit, but made sure the observational constraints encompassed the results from the fit.  The error on the mass was increased by $50\,\%$ to account for possible uncertainties resulting from the limited time span over which $\beta$ Pic b has been observed compared to the estimated orbital period (thus affecting the stellar mass estimate).  We excluded solutions with masses or radii more than $3\sigma$ away from the estimated values in order to avoid having the classic constraints drowned out by the seismic ones.  Also, we preferred the $\vsini$ value from \citet{koen2003b} over that of \citet{royer2002} due to a slightly larger resolution and signal-to-noise ratio in the spectroscopic observations, although we do note that the two values are within $2\sigma$ of each other.  We recall that the values and errors of $\vsini$ and $\logg$ intervene when calculating the radius during the MCMC runs (see Eq.~\ref{eq:cost_R}).  These constraints are summarized in the last column of Table~\ref{tab:classic_constraints}.

\begin{figure*}
\begin{center}
\includegraphics[width=\textwidth]{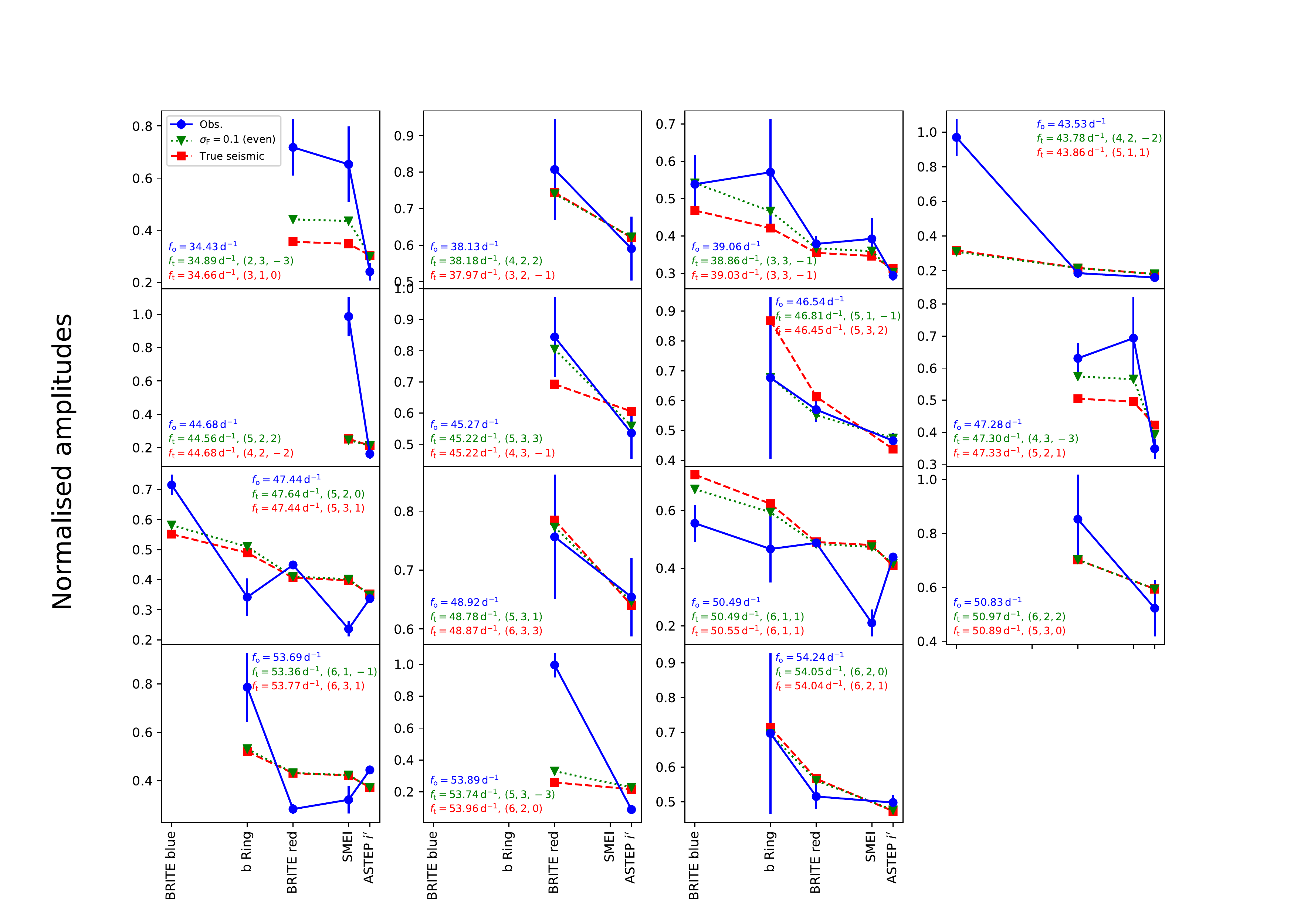} \\
\includegraphics[width=0.82\textwidth]{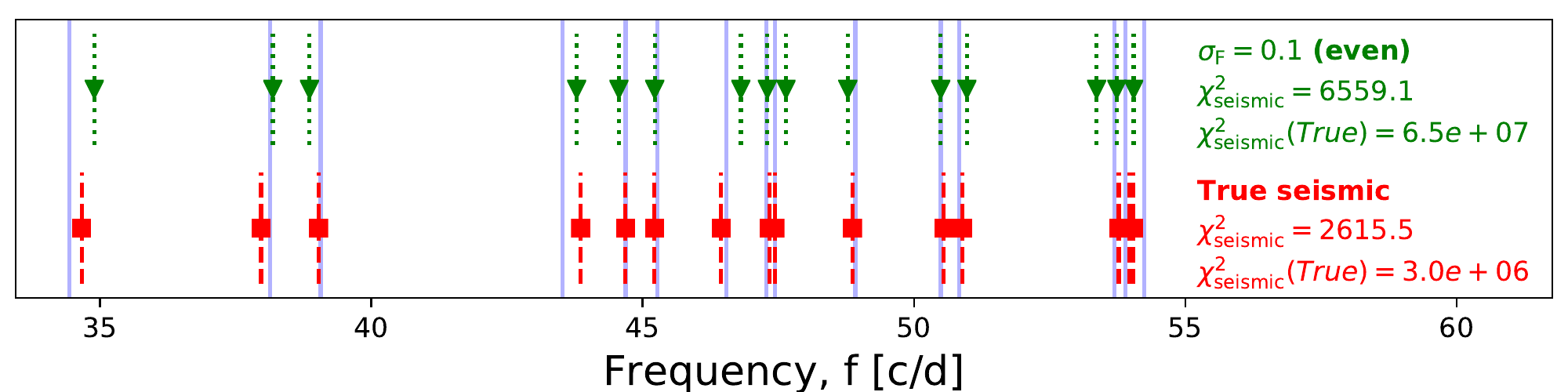}
\caption{(colour online) (Upper panel) Observed and best-fitting mode normalised amplitudes in different photometric bands using the true errors and $\sigmaF=0.1$ on the frequencies.  In the latter case, only even modes are retained and the inclination search interval is $[70^{\circ}, 90^{\circ}]$ rather than $[0^{\circ}, 90^{\circ}]$.The $(n,\,\ell,\,m)$ mode identifications are provided in each panel. (Lower panel) Observed vs. theoretical pulsation spectra for the two above cases. The observed frequencies are represented by the continuous vertical lines that span the plot.  In both panels, the observations are in blue and the theoretical results and annotations are colour-coded. 
$\chi^2$ calculations have 18 degrees of freedom, i.e., $\Omega / \OmegaK$, $i$, $f$ and the 15 free amplitudes. In the upper panel, $\chi^{2}_{ampl}$ for the $\sigma_F=0.1$ (even) case is 412.3, while $\chi^{2}_{ampl}$ for the true seismic case is 467.8.} 
\label{fig:amplitudes_main}
\end{center}
\end{figure*}

\begin{figure*}
\begin{center}
\includegraphics[width=0.8\textwidth]{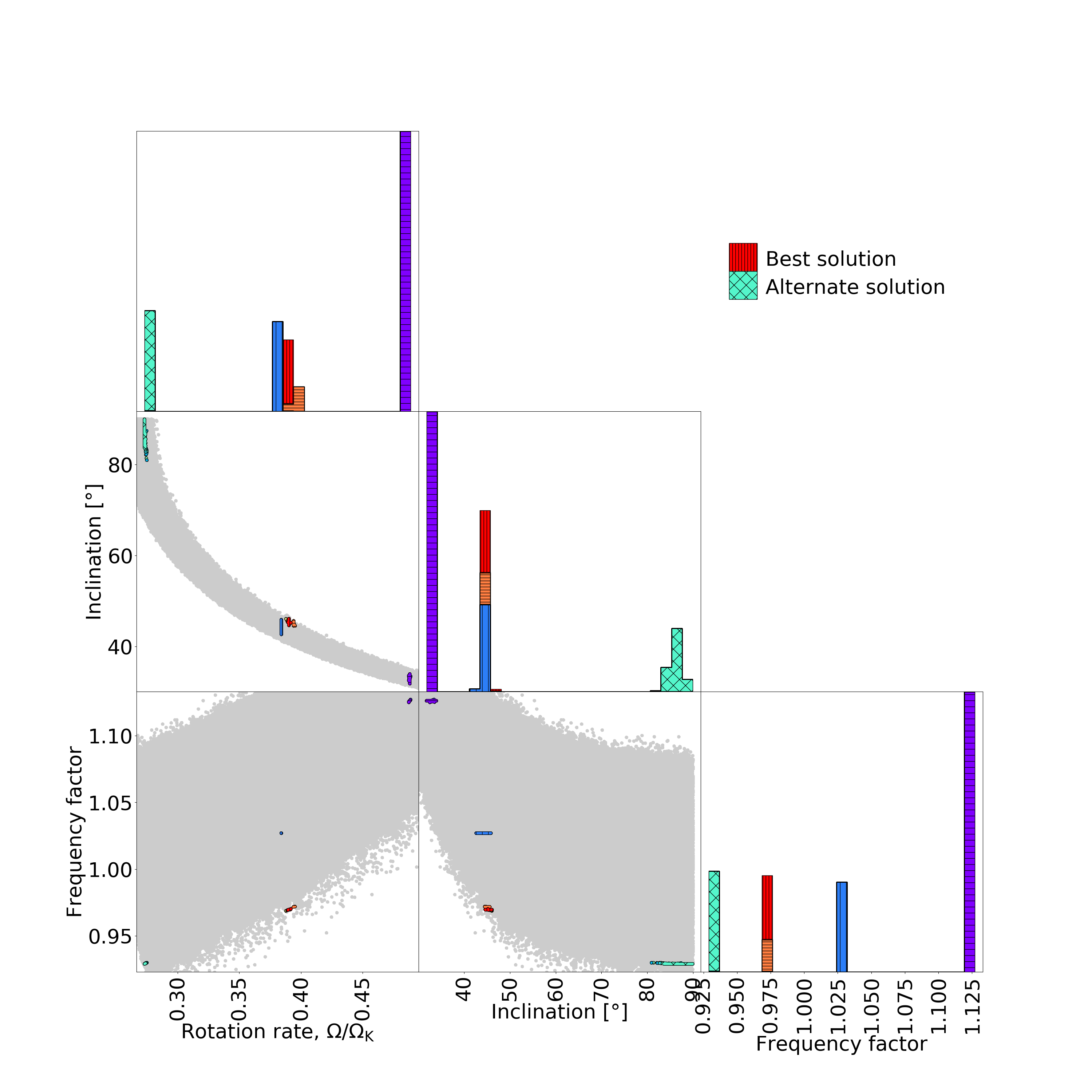}
\caption{(colour online) Distribution of solutions resulting from the seismic contraints (with the true observational errors), normalised amplitudes, and observed values of $\logg$ and $\vsini$.  The plots along the diagonal are histograms for single variables whereas the remaining plots are scatter plots for pairs of variables.  The plots are colour-coded according to the mode identification.  The solutions shown in red have the same $n$, $\ell$, $m$ identification as the best solution.}
\label{fig:true_seismic_triangle}
\end{center}
\end{figure*}

\begin{figure*}
\begin{center}
\includegraphics[width=0.8\textwidth]{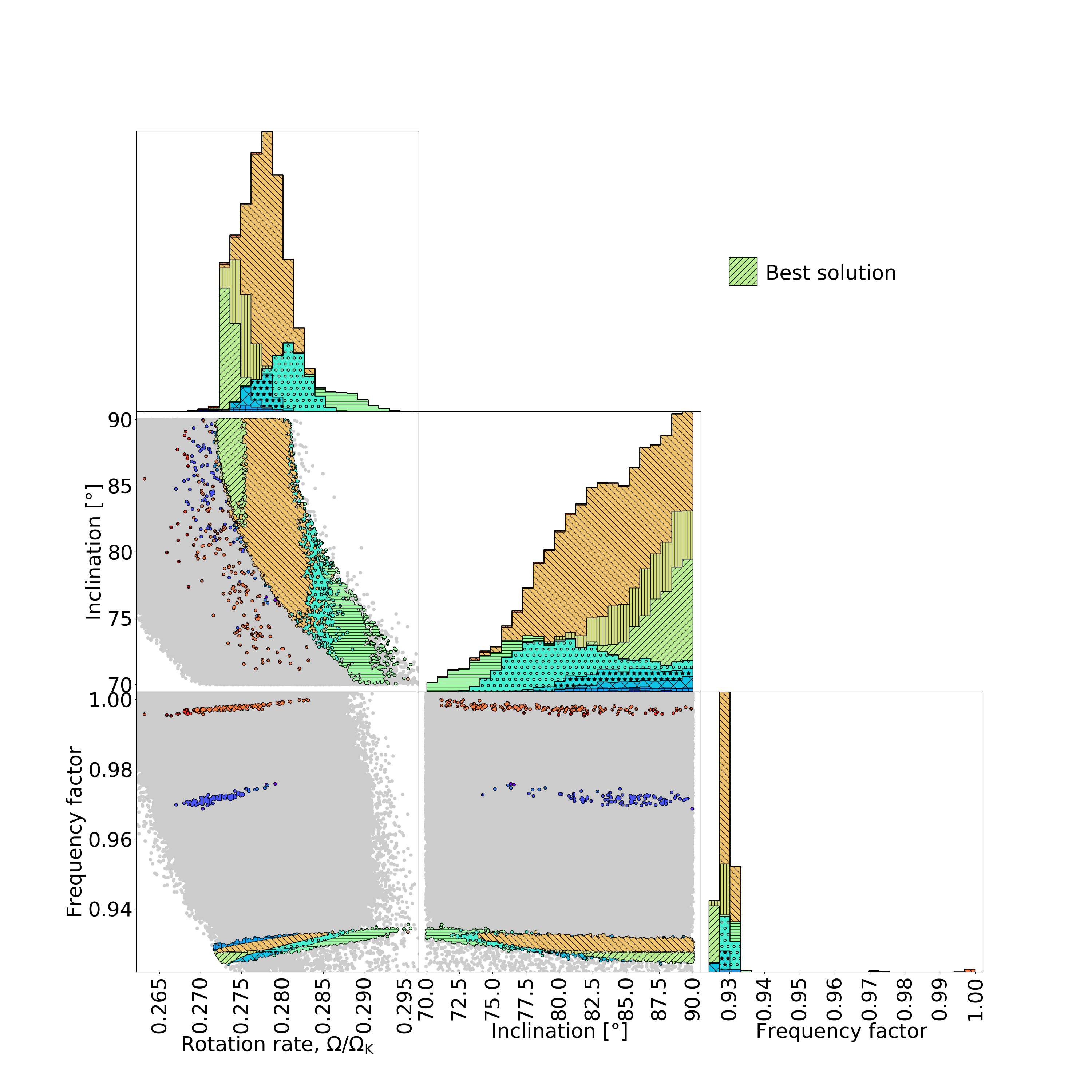}
\caption{(colour online) Same as Fig.~\ref{fig:true_seismic_triangle} except $\sigmaF=0.1$ \cd\ has been used as a frequency tolerance, the inclination has been restricted to the $[70^{\circ},90^{\circ}]$ interval and odd modes have been excluded.}
\label{fig:tol01even_triangle}
\end{center}
\end{figure*}

We first start with an MCMC run using the observational errors on the frequencies and on the amplitudes.  Figure~\ref{fig:amplitudes_main} shows the best solution\footnote{This is the solution with the lowest $\chi^2$ value.  This does not necessarily imply that its mode identification is the most represented in the sample of solutions.  Hence, it may not be the dominant colour in the triangle plots (Figs.~\ref{fig:true_seismic_triangle} and~\ref{fig:tol01even_triangle}).} found from the run.  The upper panel compares observed and theoretical amplitude in different photometric bands, whereas the lower panel compares the resultant spectra.  A fairly good fit to the spectrum is achieved but at the expense of the normalised amplitudes.  Indeed, the extremely small errors on the frequencies mean that the normalised amplitudes are hardly taken into account.  The parameters of this fit are given in the first row of Table~\ref{tab:best_solutions} and the corresponding mode identifications are provided both in Fig.~\ref{fig:amplitudes_main} and the second column of Table~\ref{tab:ids}.

In order to get a more complete picture of the posterior probability distribution resulting from the above constraints, we show a triangle plot of the distribution of solutions obtained via the MCMC run in Fig.~\ref{fig:true_seismic_triangle}.  This figure contains scatter plots for pairs of variables and histograms for single variables.  These are colour-coded according to the mode identification in the solution.  The MCMC run produced several isolated groups of solutions in the parameter space, associated with different mode identifications.  Accordingly, providing errors which cover all of these solutions is not meaningful -- instead, in what follows, we provide errors along with the statistical averages for the group of solutions with the same identification as the best solution, i.e. the solution which minimizes our cost function.  These are provided on the line below the best solution in Table~\ref{tab:best_solutions} for each configuration.  Nonetheless, some of the other groups of solutions can also be quite important.  For instance, the turquoise group contains a number of solutions with inclinations ranging from $83$ to $90^{\circ}$, which seems more realistic given the orientation of the orbit of $\beta$ Pic b and the disk around the star.  The statistical averages and standard deviations are provided on the third line of Table~7 (excluding the line with the headers).  The corresponding mode identifications are provided in the third column of Table~\ref{tab:ids}.

\begin{table*}[htbp]
\begin{center}
\caption{Best solutions obtained for different MCMC runs.\label{tab:best_solutions}}
\begin{scriptsize}
\begin{tabular}{ccccccccc}
\hline
MCMC run &
$\Omega/\OmegaK$  &
$i$ (in $^{\circ}$) &
$f$ (in d$^{-1}$) &
$R$ (in $R_{\odot}$) &
$M$ (in $M_{\odot}$) &
$\chi^2_{\mathrm{ampl.}}$ &
$\chi^2_{\mathrm{seismic}}$ &
$\chi^2_{\mathrm{seismic}}(\mathrm{true})$ \\
\hline
True seismic & 0.3903 & 45.3 & 0.96960 & 1.638 & 1.718 & 467.8 & 2615 & $3.0 \times 10^6$ \\
             & $0.3903 \pm 0.0002$ & $45.3 \pm 0.2$ & $0.96960 \pm 0.00005$ & $1.638 \pm 0.006$ & $1.719 \pm 0.020$ & & & \\
             & $0.2732 \pm 0.0001$ & $86.1 \pm 1.3$ & $0.92894 \pm 0.00006$ & $1.652 \pm 0.002$ & $1.793 \pm 0.007$ & & & \\
$\sigmaF = 0.01$ & 0.3977 & 42.9 & 1.0241 & 1.595 & 1.755 & 468.8 & 1017 & $1.2 \times 10^7$ \\
                 & $0.3977 \pm 0.0002$ & $42.9 \pm 0.5$ & $1.0241 \pm 0.0001$ & $1.595 \pm 0.015$ & $1.756 \pm 0.049$ & & &  \\
$\sigmaF = 0.05$ & 0.2625 & 80.4 & 1.0656 & 1.505 & 1.800 & 412.7 & 2257 & $7.8 \times 10^7$ \\
                 & $0.2624 \pm 0.0011$ & $82.3 \pm 3.7$ & $1.0656 \pm 0.0004$ & $1.501 \pm 0.011$ & $1.786 \pm 0.041$ & & & \\
$\sigmaF = 0.05$ (2017/18) & 0.2599 & 86.0 & 1.0630 & 1.505 & 1.794 & 508.8 & 2017 & $2.8 \times 10^7$ \\
                           & $0.2601 \pm 0.0011$ & $84.8 \pm 3.2$ & $1.0630 \pm 0.0005$ & $1.509 \pm 0.009$ & $1.809 \pm 0.032$ \\
$\sigmaF = 0.05$ (even) & 0.2742 & 76.5 & 0.9978 & 1.570 & 1.776 & 449.1 & 4723 & $1.3 \times 10^8$ \\
                        & $0.2736 \pm 0.0015$ & $78.6 \pm 3.6$ & $0.9976 \pm 0.0003$ & $1.565 \pm 0.015$ & $1.758 \pm 0.052$ & & & \\
$\sigmaF = 0.05$ (even, 2017/18) & 0.2799 & 79.8 & 0.9298 & 1.636 & 1.737 & 528.9 & 7211 & $9.0 \times 10^7$ \\
                                 & $0.2801 \pm 0.0013$ & $81.5 \pm 3.6$ & $0.9298 \pm 0.0005$ & $1.631 \pm 0.012$ & $1.720 \pm 0.039$ & & & \\
$\sigmaF = 0.1$ (even) & 0.2723 & 89.9 & 0.9258 & 1.658 & 1.802 & 412.3 & 6559 & $6.5 \times 10^7$ \\
                       & $0.2735 \pm 0.0007$ & $87.8 \pm 1.6$ & $0.9262 \pm 0.0004$ & $1.653 \pm 0.004$ & $1.785 \pm 0.013$ & & & \\
$\sigmaF = 0.1$ (even, 2017/18) & 0.2723 &  89.2 & 0.9258 & 1.658 & 1.802 & 515.9 & 8049 & $6.2 \times 10^7$ \\
                               & $0.2735 \pm 0.0007$ & $87.6 \pm 1.6$ & $0.9262 \pm 0.0004$ & $1.653 \pm 0.004$ & $1.785 \pm 0.013$ & & & \\
\hline
\end{tabular}
\end{scriptsize}
\end{center}
\tablefoot{Two lines are provided for each configuration apart from the first configuration.  The upper line shows the best solution (i.e. which minimizes the cost function).  The $\chi^2_{\mathrm{seismic}}$ values given in the second-last column have all been calculated using a frequency tolerance $\sigmaF=0.01$ d$^{-1}$ even if the MCMC run may have used a different value of $\sigmaF$.  The second line gives the statistical averages and standard deviations of all of the solutions \textit{with the same mode identification} as the best solution.   As such, these uncertainties do not account for the dispersion resulting from other identifications, or from systematic effects which may occur as a result of limitations in the models.  The third line for the first configuration corresponds to a near equator-on set of secondary solutions obtained for the first MCMC run.}
\end{table*}

\begin{table*}
\caption{Mode identifications in the form $(n,\ell,m)$ for best and alternate configurations. \label{tab:ids}}
\begin{tabular}{lccccccccc}
\hline
No. & \multicolumn{2}{c}{\dotfill True seismic \dotfill} & $\sigmaF=0.01$ & $\sigmaF=0.05$ &  $\sigmaF=0.05$ &  $\sigmaF=0.05$ &  $\sigmaF=0.05$ &  $\sigmaF=0.1$ & $\sigmaF=0.1$ \\
\#   &  Best &   Alternate   &  &  & (2017/18) & (even) & (even, 2017/18) & (even) & (even, 2017/18) \\
\hline
F1 & $(3,1,0)$ &   $(3,2,1)$ &   $(2,2,-2)$ &  $(1,3,-3)$ &  $(2,3,1)$ &   $(3,2,2)$ &   $(3,1,-1)$ &  $(2,3,-3)$ &  $(2,3,-3)$ \\
F2 & $(3,2,-1)$ &  $(4,2,2)$ &   $(4,3,3)$ &   $(3,3,3)$ &   $(3,3,3)$ &   $(3,2,0)$ &   $(4,2,2)$ &   $(4,2,2)$ &   $(4,2,2)$ \\
F3 & $(3,3,-1)$ &  $(4,3,3)$ &   $(4,1,1)$ &   $(3,1,-1)$ &  $(3,1,-1)$ &  $(3,3,1)$ &   $(3,3,-1)$ &  $(3,3,-1)$ &  $(3,3,-1)$ \\
F4 & $(5,1,1)$ &   $(4,2,-1)$ &  $(4,1,-1)$ &  $(3,3,0)$ &   $(3,3,0)$ &   $(4,1,-1)$ &  $(5,0,0)$ &   $(4,2,-2)$ &  $(4,2,-2)$ \\
Fc & $(4,2,-2)$ &  $(5,2,2)$ &   $(5,3,3)$ &   $(3,3,-1)$ &  $(3,3,-1)$ &  $(4,2,0)$ &   $(5,2,2)$ &   $(5,2,2)$ &   $(5,2,2)$ \\
F5 & $(4,3,-1)$ &  $(5,3,3)$ &   $(5,2,2)$ &   $(4,3,3)$ &   $(4,3,3)$ &   $(4,3,1)$ &   $(4,3,-1)$ &  $(5,3,3)$ &   $(5,3,3)$ \\
F6 & $(5,3,2)$ &   $(5,1,0)$ &   $(4,3,0)$ &   $(4,1,-1)$ &  $(4,1,-1)$ &  $(5,0,0)$ &   $(5,1,-1)$ &  $(5,1,-1)$ &  $(5,1,-1)$ \\
F7 & $(5,2,1)$ &   $(5,2,1)$ &   $(4,2,-2)$ &  $(3,3,-3)$ &  $(4,3,2)$ &   $(4,2,-2)$ &  $(4,3,-3)$ &  $(4,3,-3)$ &  $(4,3,-3)$ \\
F8 & $(5,3,1)$ &   $(4,3,-3)$ &  $(4,3,-1)$ &  $(4,2,0)$ &   $(4,2,0)$ &   $(5,1,1)$ &   $(5,2,0)$ &   $(5,2,0)$ &   $(5,2,0)$ \\
F9 & $(6,3,3)$ &   $(5,3,1)$ &   $(5,3,2)$ &   $(4,3,1)$ &   $(4,3,1)$ &   $(4,3,-1)$ &  $(5,3,1)$ &   $(5,3,1)$ &   $(5,3,1)$ \\
F11 & $(6,1,1)$ &   $(5,2,-2)$ &  $(5,1,-1)$ &  $(4,2,-2)$ &  $(4,2,-2)$ &  $(5,1,-1)$ &  $(6,1,1)$ &   $(6,1,1)$ &   $(6,1,1)$ \\
F12 & $(5,3,0)$ &   $(5,3,0)$ &   $(5,1,0)$ &   $(5,1,1)$ &   $(4,3,0)$ &   $(4,3,-3)$ &  $(6,2,2)$ &   $(6,2,2)$ &   $(6,2,2)$ \\
F13 & $(6,3,1)$ &   $(6,2,1)$ &   $(5,3,0)$ &   $(5,1,0)$ &   $(4,3,-2)$ &  $(6,0,0)$ &   $\mathbf{(5,3,-3)}$ &  $(6,1,-1)$ &  $\mathbf{(5,3,-3)}$ \\
F14 & $(6,2,0)$ &   $(5,3,-3)$ &  $(6,0,0)$ &   $(4,3,-2)$ &  $(5,1,-1)$ &  $(5,2,-2)$ &  $\mathbf{(6,1,-1)}$ &  $(5,3,-3)$ &  $\mathbf{(6,1,-1)}$ \\
F15 & $(6,2,1)$ &   $(6,2,0)$ &   $(5,3,-1)$ &  $(5,2,1)$ &   $(5,2,1)$ &   $(6,1,1)$ &   $(6,2,0)$ &   $(6,2,0)$ &   $(6,2,0)$ \\
\hline
\end{tabular}
\tablefoot{The boldfaced mode identifications in columns 8 and 10 correspond to modes for which the theoretical frequencies are in the wrong order compared to observations (i.e. the two modes are swapped), likely as a result of favoring normalised amplitudes over frequencies.}
\end{table*}

In order to increase the weight of the observed amplitudes in the fit, we carried out other MCMC runs using a uniform frequency tolerance, denoted $\sigmaF$.  This frequency tolerance acts as a trade-off parameter between fitting the spectrum and matching the amplitudes.  If a small value is chosen for $\sigmaF$, the MCMC algorithm will favor solutions where the frequencies are a good match to the observations, but the amplitudes will be a poor fit.  Conversely, large values of $\sigmaF$ lead to the opposite behavior.  Figure~\ref{fig:amplitudes_appendix} in Appendix~\ref{sect.extra_astero} shows the solutions obtained for $\sigmaF=0.01$\cd\ and $0.05$\cd\, respectively.  The corresponding identifications are provided in these figures as well as in Table~\ref{tab:ids}.  The resultant parameters are given in lines 4 to 7 of Table~\ref{tab:best_solutions} (excluding the header line).  The different choices of $\sigmaF$ have lead to rather different solutions but which both match the observations fairly well.  Moreover, the solution for $\sigmaF=0.05$\cd\, is fairly close to equator-on.  This may be realistic for $\beta$ Pic due to the nearly edge-on configuration of its disk and the planet's orbit, and the lack of plausible mechanisms able to misalign the system.  Nonetheless, this solution includes a few odd modes, i.e. modes which are antisymmetric with respect to the equator, as can be seen in Table~\ref{tab:ids} (by calculating the parity of $\ell+m$).  This seems somewhat unrealistic since such modes should cancel out for an equator-on configuration and thus not be visible.  Accordingly, we carried out an MCMC run for $\sigmaF=0.05$\cd\, including only even modes and restricting the inclination to values between $70^{\circ}$ and $90^{\circ}$ (rather than $0^{\circ}$ to $90^{\circ}$).  The corresponding solution is also shown in Fig.~\ref{fig:amplitudes_appendix}, and the resultant parameters given in Table~\ref{tab:best_solutions}.  Although this solution has fairly similar parameter values for $\Omega/\OmegaK$ and $i$ as the previous solution including even and odd modes, this solution corresponds to a completely different set of mode identifications.  Furthermore, the inclination $i=76.8^{\circ}$ is not entirely satisfactory as it is sufficiently far from equator-on to invalidate the exclusion of odd modes.  

A further MCMC run is carried out using $\sigmaF = 0.1$\cd\ in search of solutions which are closer to equator-on.  The best solution is shown in Fig.~\ref{fig:amplitudes_main} and the corresponding parameters given in Table~\ref{tab:best_solutions}. This solution is much closer to equator on.  In fact, the statistical average value for the inclination provided in Table~\ref{tab:best_solutions} is biased by the fact that the inclination range is bounded by $90^{\circ}$ thus artificially leading to a lower average. This best solution contains several modes with a similar identification as was obtained for $\sigmaF=0.05$\cd\ with even and odd modes, apart from an offset of $1$ on the radial order.  Finally, Figure~\ref{fig:tol01even_triangle} provides a colour-coded triangle plot showing the distribution of solutions in parameter space.  A diversity of identifications are obtained.  Furthermore, the group associated with the best solution (light green-yellow) 
clearly peaks at $90^{\circ}$, thus favoring an equator-on configuration.

Another factor to be taken into account is the fact that a couple of the amplitudes change significantly in the BRITE R band between two observational runs.  Indeed, the F11 mode at 50.49\cd\ has an amplitude of 0.84 mmag in 2016/17 and increases to 1.06 mmag in 2017/18.  Likewise, the F13 mode at 53.69\cd\ has an amplitude of 0.40 mmag in 2016/17 in BRITE R and increases to 1.27 mmag in 2017/18.  The fits carried out so far were primarily based on the 2016/17 data.  We therefore carried out a few more MCMC runs using the 2017/18 data instead.  The resultant best solutions as well as relevant statistical averages and standard deviations are listed in Table~\ref{tab:best_solutions}.  The corresponding mode identifications are provided in Table~\ref{tab:ids}.  The choice of observing season does have some impact on the values of $\Omega/\OmegaK$, $i$, and $f$, especially for $\sigmaF=0.05$\cd\ using only even modes.  The mode identification is completely modified for this particular case, but is hardly affected for $\sigmaF=0.1$\cd\ with even modes.  

The fact that modifying the amplitudes of two modes can affect the entire mode identification is not entirely surprising since the MCMC algorithm is optimizing the fit to all of the modes simultaneously.  Also, in almost all cases, the seismic and amplitude-related values of $\chi^2$ are degraded.  This shows that the models clearly provide a better match to the $2016/17$ run over the $2017/18$ run.  One possible explanation for this worse fit is the fact that the observations in the various photometric bands are not simultaneous and are for the most part more representative of the $2016/17$ time period.  Ideally, amplitude changes should proportionally be the same in the different bands over similar time periods, thus canceling out when calculating normalised amplitudes given that these only depend on mode geometry.

Overall, our favored solution is the one obtained for $\sigmaF=0.1$\cd, using only even modes.  Indeed, this solution seems to be the most coherent in terms of stellar inclination, given the measured inclination of the planetary orbit and the circumstellar disk of 88.81$\pm 0.12^{\circ}$ \citep{wang2016}.  It also leads to the best fit with the normalised amplitudes.  However, the price to pay is a relatively high $\chi^2$ value for the seismic component (although the theoretical frequencies come in the same order as the observed ones).  Possible causes for this significant difference in the frequencies include shortcomings in the stellar models.  In particular, the models are uniformly rotating.  This does not seem very realistic because baroclinic effects are expected to lead to differential rotation as shown in more realistic models based on the ESTER code \citep{espinosa2013,rieutord2016}.  This in turn will affect rotational splittings \citep{reese2009}, thus modifying the frequencies of non-axisymmetric modes.  Also, the differences between the fitted and observed mode amplitudes still remain relatively high even for the most favorable solution.  Possible causes for this include the use of pseudo non-adiabatic mode visibilities rather than visibilities based on fully non-adiabatic calculations.  

However, at this point it is not possible to obtain reliable fully non-adiabatic calculations.  Indeed, one would need rapidly rotating models which solve the energy conservation equation in a consistent way -- this is currently only achieved in the ESTER code.  However, the ESTER code does not currently model sub-surface convective envelopes which is expected to be relevant in a star of this mass.  Another limitation in the visibility calculations is the fact that they do not rely on realistic model atmospheres but rather on blackbody spectra as was pointed out earlier.  Finally, another factor which needs to be considered is the set of modes used in the identification.  Indeed, we restricted ourselves to modes with $\ell$ between $0$ and $3$.  However, visibility calculations at rapid rotation rates suggest that higher $\ell$ modes may become more visible \citep[e.g.][]{lignieres2009}.  Carrying out fits with higher $\ell$ values does lead, as expected, to better fits with $\chi^2_{\mathrm{ampl}}$ below $400$ and/or $\chi^2_{\mathrm{seismic}}$ below $1000$ in some cases.  Hence, an alternate approach may be to select modes based on their visibilities at the relevant rotation rate, rather than using a predefined set of modes as was done above.  Nonetheless, non-linear mode coupling and saturation mechanisms may lead to intrinsic mode amplitudes that alter which modes are actually observed compared to what would be expected from geometric visibility factors.

\section{Discussion and Conclusion}

The exoplanet host star \bpic\, was already known to show p-mode pulsations of \dsct-type \citep{koen2003a,koen2003b,mekarnia17}. 
As observations with the {\it Kepler} space telescope \citep{borucki2010} revealed that many \dsct\ stars show both, p- and g-modes \citep{uytterhoeven2011}, we also investigated the presence of g-modes in our \bpic\ data sets which would be expected to lie in the frequency range between 0.3 and 3\,\cd\ \citep{aerts2010}. 
BRITE-Constellation observations are known to be in particular sensitive to frequencies in this range \citep[e.g.,][]{baade2018,tahina2018} due to the satellites' observing strategy \citep{weiss2014}.

Using both, our best data set alone (i.e., the BRITE R filter observations of 2016/17 which reach the lowest residual noise level) and a combination of three seasons of BRITE R filter observations, we do not find evidence for the presence of g-modes down to a residual noise level of 36\,ppm. It is therefore evident that if g-modes exist in \bpic, they must possess even lower amplitudes that remain undetected in the  data sets analyzed here.  

Our 15 identified pulsation frequencies correspond to the 14 highest amplitude frequencies in \citet{mekarnia17} which are numbered f$_1$ to f$_{14}$ and their frequency f$_{23}$. As the residual noise level of 9.45\,ppm is significantly lower in \citet{mekarnia17} compared to our best data set (i.e., the BRITE R filter observations obtained in 2016/17) with a residual noise level of 40\,ppm, not all the pulsation frequencies reported earlier are identified to be significant in our analysis.

We used the amplitudes of 15 \dsct-type p-mode frequencies 
detected in up to five different passbands - BRITE B \& R, SMEI, bRing and ASTEP $i^{\prime}$ - to calculate normalised amplitudes for an asteroseismic study of $\beta$ Pic and an identification of its pulsation modes. 
This analysis was complicated by the fact that two pulsation frequencies show a clear variability in our time series observations: The frequency F13 at 53.6917\cd\ first decreases between the 2015 and 2016/17 observations and then increases again between 2016/17 and 2018; the amplitude of frequency F11 at 50.4921\cd\ constantly increases in the observations obtained between 2015 and 2018. This behaviour could consistently be found in the BRITE-Constellation and the bRing data and confirms earlier reports by \citet{mekarnia17}. All other pulsation modes have stable amplitudes within the observational errors.

In general, the variability of certain amplitudes should proportionally be the same in different bands over similar periods of time; thus amplitude variability should not impact the calculations of  normalised amplitudes given that these only depend on mode geometry. In the present case, the observations obtained by BRITE-Constellation, bRing and ASTEP were not taken exactly simultaneously, but with overlapping periods of time, and the SMEI photometry was obtained years before. Hence, the variable amplitudes of the two modes affect the observed normalised amplitudes and thus the identification of the pulsation modes.

For our asteroseismic interpretation of the  normalised amplitudes we used the most precise BRITE R filter data set (i.e., the $\sim$224-day long BRITE R filter observations obtained in 2016/17 with a residual noise level of 40\,ppm), the simultaneous BRITE B filter observations of 2016/17, the overall amplitudes obtained from the bRing and SMEI data and the previously published ASTEP $i^{\prime}$ amplitudes.
From this, our favored solution of the asteroseismic models was obtained for a relatively large value for the frequency tolerance $\sigmaF=0.1$ \cd\ on the frequencies, only including even modes.  
This leads to the best fit to the  normalised amplitudes, while at the same time getting a near equator-on inclination of $i=89.1^{\circ}$, which is in agreement with our expectations based on the orbital inclination of $\beta$ Pic b as well as that of the circumstellar disk.
Correspondingly, the pulsation modes were identified as three $\ell=1$, six $\ell=2$ and six $\ell=3$ modes. 
Our preferred model also yields a rotation rate of $\sim$27\% of Keplerian breakup velocity, a radius of $1.497 \pm 0.025$ \Rsun\, and a mass of $1.797 \pm 0.035$ \Msun\, corresponding to an error of $\sim$2\% in stellar mass and less than 2\% in stellar radius.
These errors do not account for uncertainties in the models and for errors resulting from an erroneous mode identification. The fact that the difference between the observations and the theory remains high implies that the model errors could be quite significant.
Hence, although the errors on mass and radius of \bpic\, are quite small, they only account for a small part of the true error. 
The choice of observing season only has a limited impact on the values of $\Omega / \OmegaK$, $i$, and $f$ when assuming a relatively large observational frequency tolerance $\sigmaF=0.1$ \cd\ on the frequencies.
Finally, the choice of what set of theoretical modes should be considered when fitting the observations remains an open question.

Our analysis yields an independent and more accurate determination of the stellar parameters based on the combination of classic constraints with the pulsational properties of $\beta$ Pictoris derived in multiple passbands. We illustrate that adding seismic constraints considerably reduces the set of acceptable theoretical models, hence, resulting in higher precision. 

Mode identification in rapidly rotating $\delta$ Scuti stars is one of the outstanding problems in stellar physics \citep[e.g.,][]{goupil2005,deupree2011}. Our work constitutes an important step in addressing this, hence illustrates the importance of good priors on the classic quantities.  However, the seismic analysis still manages to further restrict the acceptable values for the mass and radius.
Additionally, $\beta$ Pictoris is a $\delta$ Scuti type star where most pulsation amplitudes remain stable over many years, while a few change sometimes even significantly. Hence, it is another candidate for future studies of the physical reasons of amplitude variability versus stability in $\delta$ Scuti stars. 
Our search for g-modes in our data sets was motivated by the idea to use them to probe the near-core region of $\beta$ Pictoris; the detection of g-modes would allow us to investigate differential rotation in the stellar interiors using prograde and retrograde pulsation modes \citep[e.g.,][]{zwintz2017} and study the angular momentum distribution \citep[e.g.,][]{aerts2017}.

The absence of a magnetic field (i.e., if there is a magnetic field, its strength has to be lower than 300 Gauss) might also be a crucial factor to be considered when studying the exoplanetary system and circumstellar disk around $\beta$ Pictoris.

\begin{acknowledgements}
The authors thank Marc-Antoine Dupret for results from the MAD code which allowed us to calculate pseudo non-adiabatic mode visibilities. We thank P. Kervella and F. Royer for helpful discussions and clarifications concerning some of the constraints on $\beta$ Pictoris.  
We also express our gratitude to our referee, Dietrich Baade, for his very constructive comments.
This work has made use of data from the European Space Agency (ESA) mission
{\it Gaia} (\url{https://www.cosmos.esa.int/gaia}), processed by the {\it Gaia}
Data Processing and Analysis Consortium (DPAC,
\url{https://www.cosmos.esa.int/web/gaia/dpac/consortium}). Funding for the DPAC
has been provided by national institutions, in particular the institutions
participating in the {\it Gaia} Multilateral Agreement.
K.Z. acknowledges support by the Austrian Fonds zur F\"orderung der wissenschaftlichen Forschung (FWF, project V431-NBL) and the Austrian Space Application Programme (ASAP) of the Austrian Research Promotion Agency (FFG). D.R.R. acknowledges the support of the French Agence Nationale de la Recherche (ANR) to the ESRR project under grant ANR-16-CE31-0007 as well as financial support from the Programme National de Physique Stellaire (PNPS) of the CNRS/INSU co-funded by the CEA and the CNES.
A.Pi. acknowledges support from the NCN grant 2016/21/B/ST9/01126.
APo was responsible for image processing and automation of photometric routines for the data registered by the BRITE nano-satellite constellation, and was supported by the statutory activities grant BK/200/RAU1/2018 t.3. 
GH thanks the Polish National Center for Science (NCN) for support through grant 2015/18/A/ST9/00578. The research of S.M.R. and A.F.J.M. has been supported by the Natural Sciences and Engineering Research Council (NSERC) of Canada. GAW acknowledges Discovery Grant support from the Natural Science and Engineering Research Council (NSERC) of Canada.
MI was the recipient of an Australian Research Council Future Fellowship  (FT130100235) funded by the Australian Government. SNM is a U.S. Department of Defense SMART scholar sponsored by the U.S. Navy through SSC-LANT. Part of this research was carried out at the Jet Propulsion Laboratory, California Institute of Technology, under a contract with the National Aeronautics and Space Administration. E.E.M. and S.N.M. acknowledge support from the NASA NExSS program. The bRing observatory at Siding Springs, Australia was supported by a University of Rochester University Research Award.
\end{acknowledgements}

\object{beta Pictoris}

\bibliographystyle{aa}
\bibliography{betapic_v3.bib}

\appendix

\section{Supplementary figures for the frequency analysis of BRITE data}
\label{appendix.FA}

\begin{figure}[htb]
\includegraphics[width=0.45\textwidth]{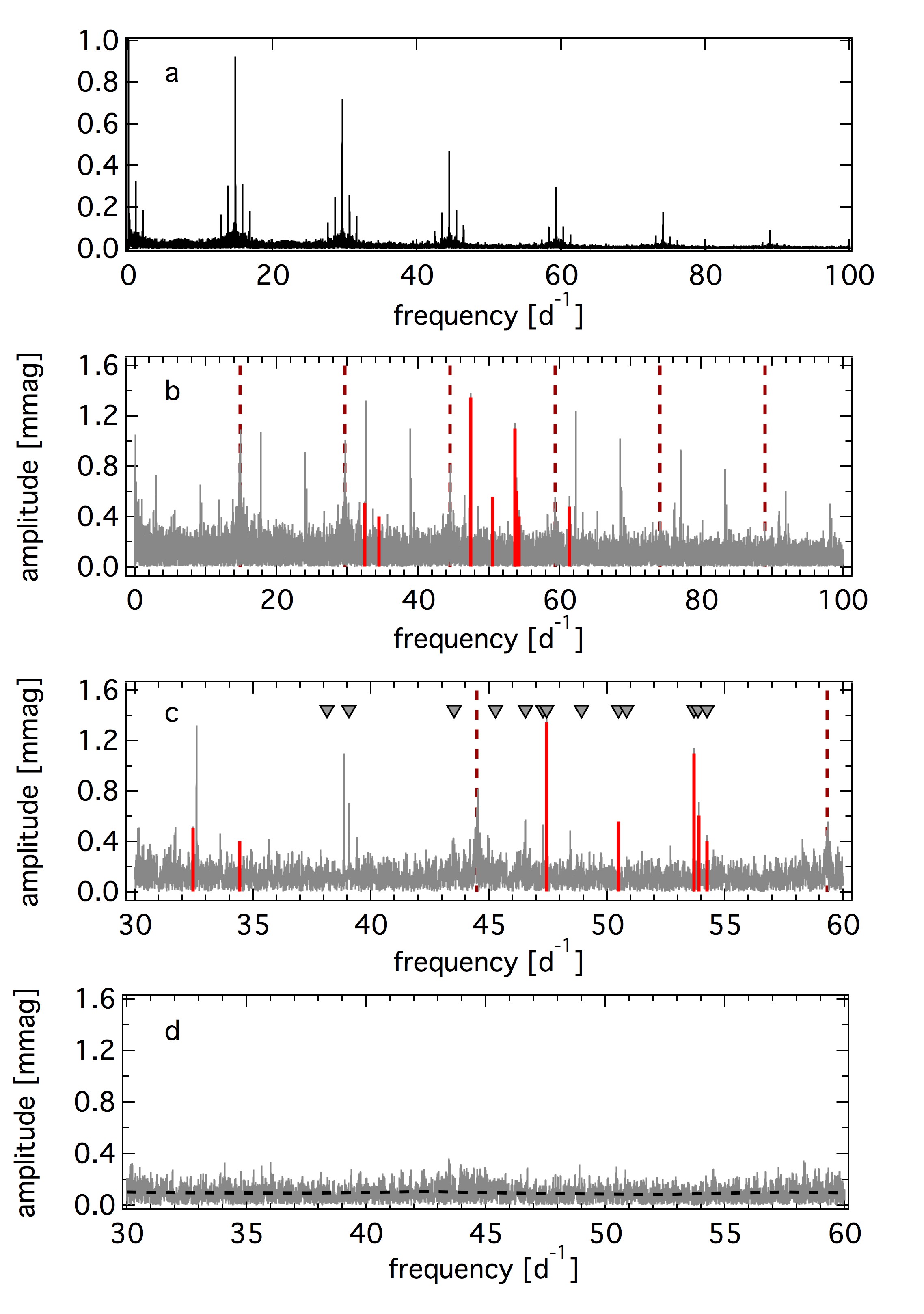}
\caption[]{Frequency analysis of the BHr 2015 data: spectral window (panel a), original amplitude spectrum from 0 to 100\cd\ (panel b), zoom into the original amplitude spectrum (panel c) and residual amplitude spectrum after prewhitening of the eight frequencies (panel d) with the residual noise level marked as horizontal dashed line. The triangles mark the frequencies found in the ASTEP data by \citet{mekarnia17}. Vertical dashed lines mark the positions of the BHr satellite's orbital frequency and its multiples.}
\label{BHr2015}
\end{figure}

Figure \ref{BHr2015} shows the amplitude spectrum of the original BHr data (grey) with the eight identified frequencies (red). The orbital frequency of BHr at 14.831\cd\ and its multiples are given as dashed dark-red lines. Note that some of the peaks with high amplitude are alias frequencies to one of the pulsation frequencies with the orbital frequency and disappear after prewhitening.

\begin{figure}[htb]
\begin{center}
\includegraphics[width=0.45\textwidth]{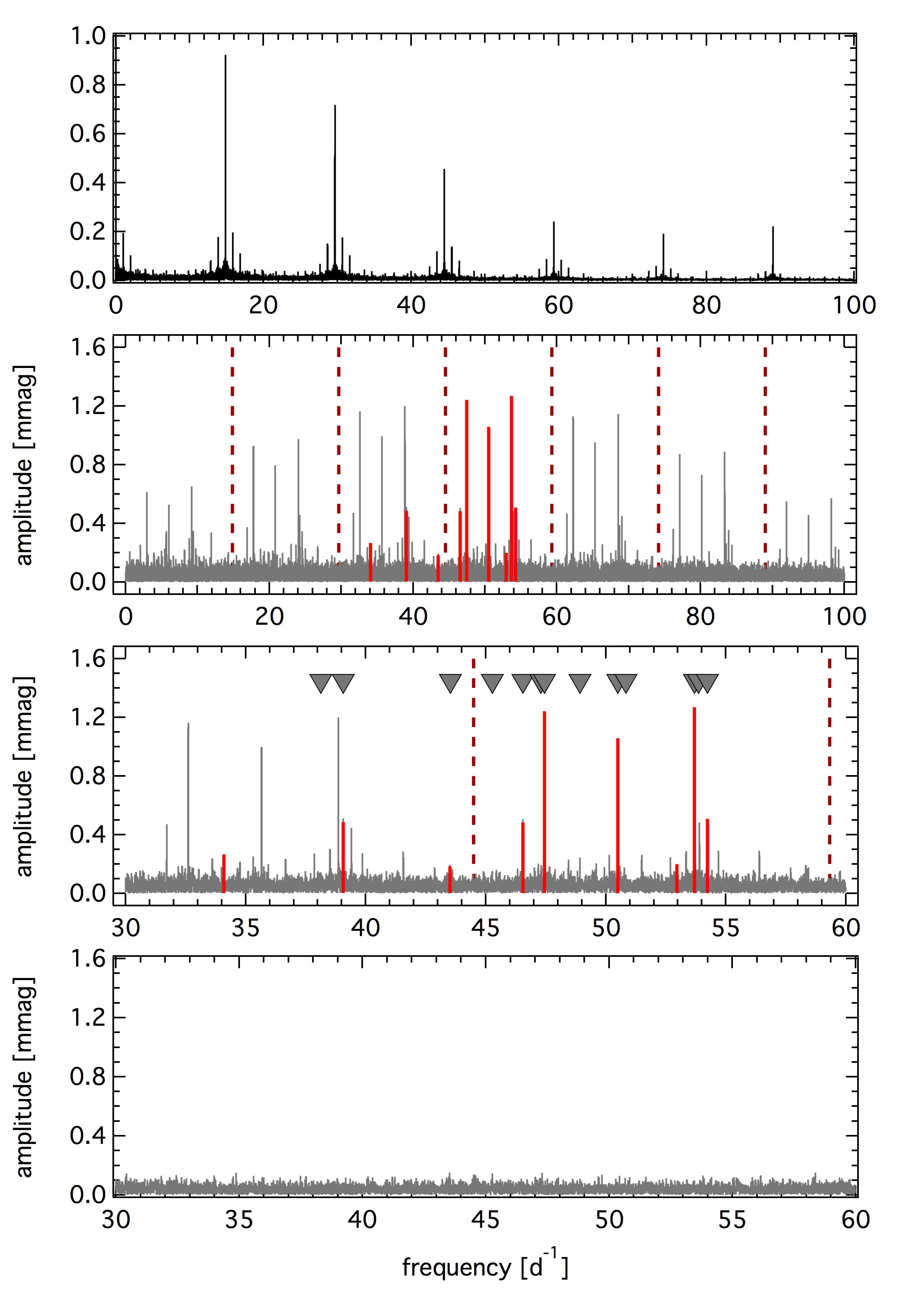}
\caption[]{Frequency analysis of the BHr 2017/2018 data: spectral window (panel a), original amplitude spectrum from 0 to 100\cd\ (panel b), zoom into the original amplitude spectrum (panel c) and residual amplitude spectrum after prewhitening of the eight frequencies (panel d). The triangles mark the frequencies found in the ASTEP data by \citet{mekarnia17}. Vertical dashed lines mark the positions of the respective satellite's orbital frequency and its multiples.}
\label{2017data}
\end{center}
\end{figure}

The amplitude spectrum using the BHr 2017/18 data is shown in Figure \ref{2017data} where again the frequencies reported by \citet{mekarnia17} are marked as grey triangles, the orbital frequency of BHr and its multiples are identified as dark red dashed lines and the spectral window is given in the top panel.

\begin{figure}[htb]
\begin{center}
\includegraphics[width=0.4\textwidth]{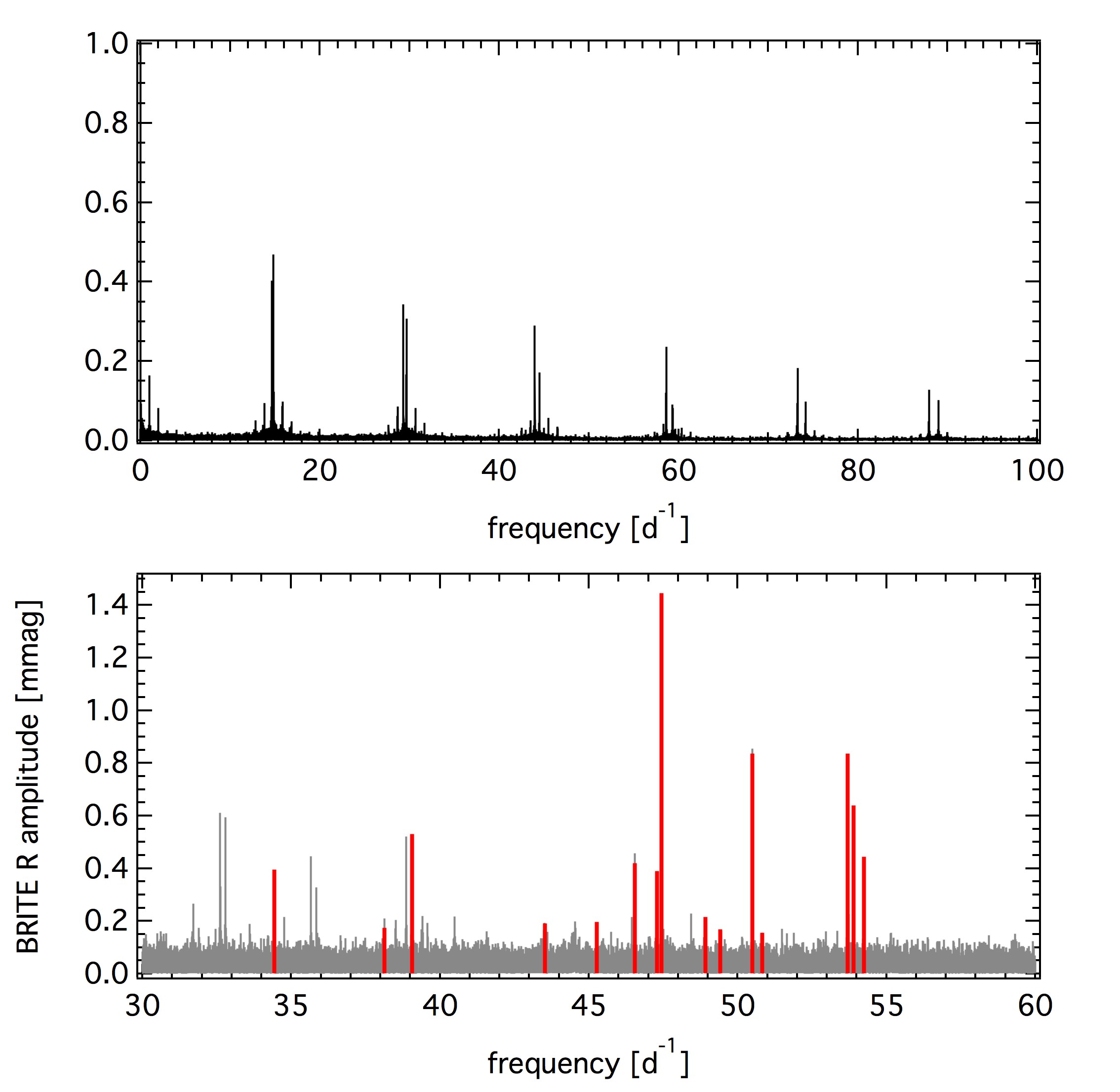}
\caption[]{Frequency analysis of the combined BRITE R filter data: spectral window (top), and original amplitude spectrum from 30 to 60\,\cd\ (bottom) with the identified pulsation frequencies marked in red.}
\label{combined_R}
\end{center}
\end{figure}

The frequency analysis of the combined BRITE R filter data yielded the same pulsation frequencies as given in Table \ref{tab:freqs}. The corresponding amplitude spectrum with the identified frequencies marked in red is shown in Figure \ref{combined_R}.

\section{Supplementary material for amplitude variability using bRing data}

Investigation of the amplitude variability using the bRing time series which consists of observations taken in South Africa and Australia. As the Australian bRing instrument had first light on December 1, 2017, corresponding to a JD of $\sim$2458091.0, we did not detect the three frequencies F8, F11, F13 in the subsets using the first about 200-days of the bRing light curve. Unfortunately the errors on the amplitudes calculated from the individual subsets are mostly too high for a reliable study of amplitude variability from this data set.

\begin{figure}
\begin{center}
\includegraphics[width=0.45\textwidth]{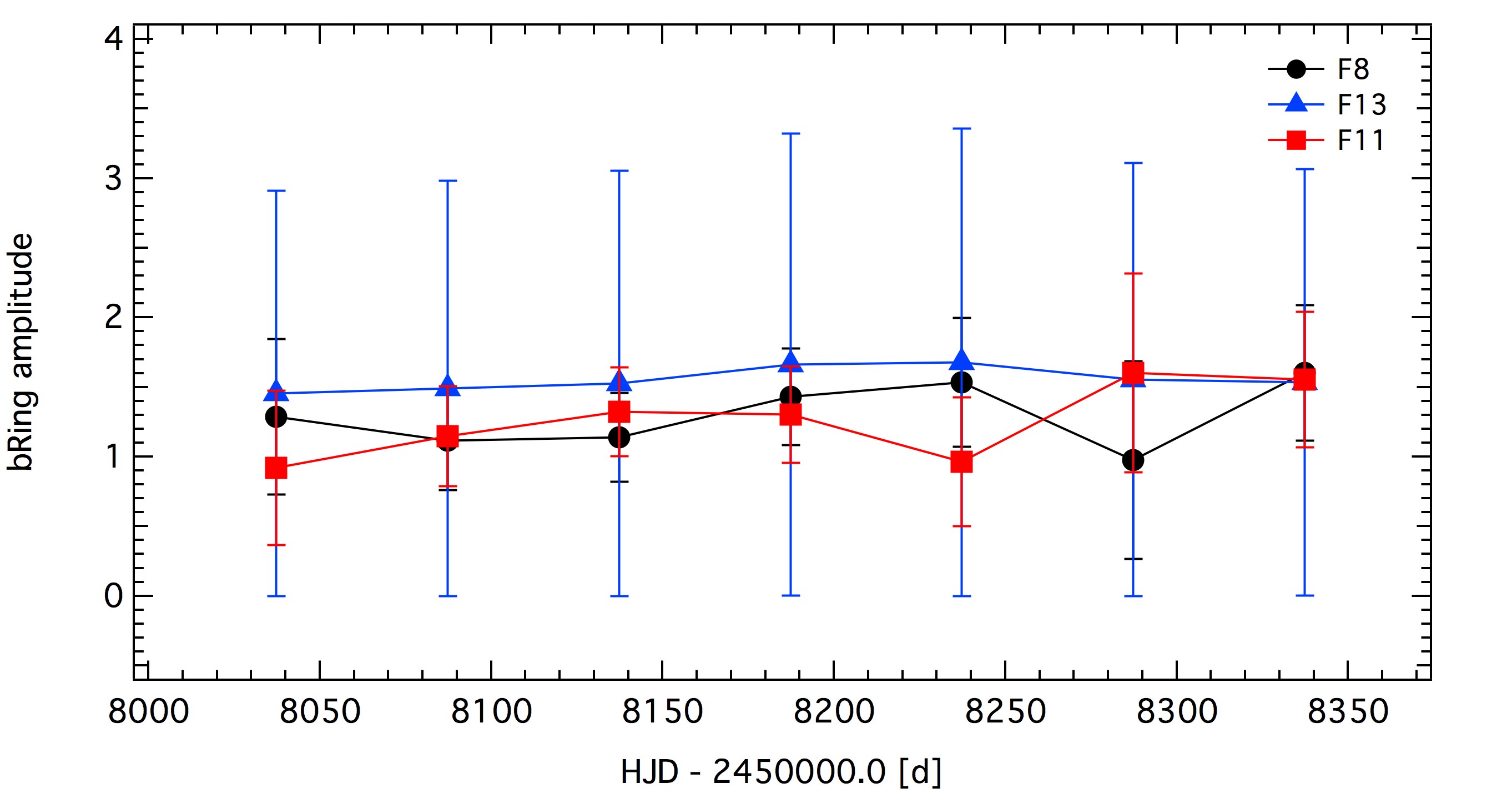}
\caption{Amplitude behaviour of the three highest-amplitude pulsation frequencies during the bRing observations calculated from 100-day subsets.}
\label{ampvar_bring}
\end{center}
\end{figure}

\section{Supplementary material for the asteroseismic interpretation}
\label{sect.extra_astero}

\begin{figure*}
\begin{center}
\includegraphics[width=\textwidth]{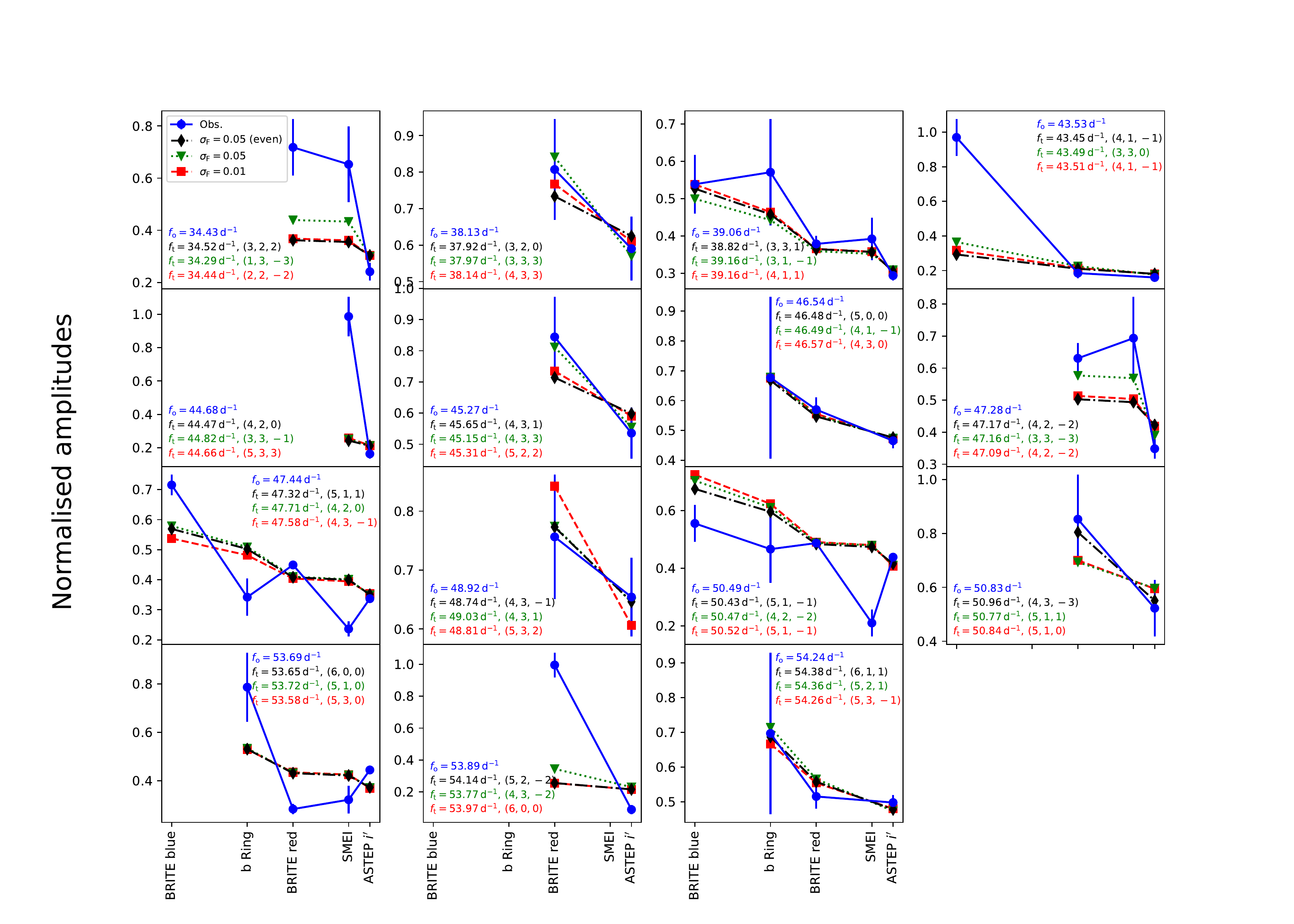} \\
\includegraphics[width=0.82\textwidth]{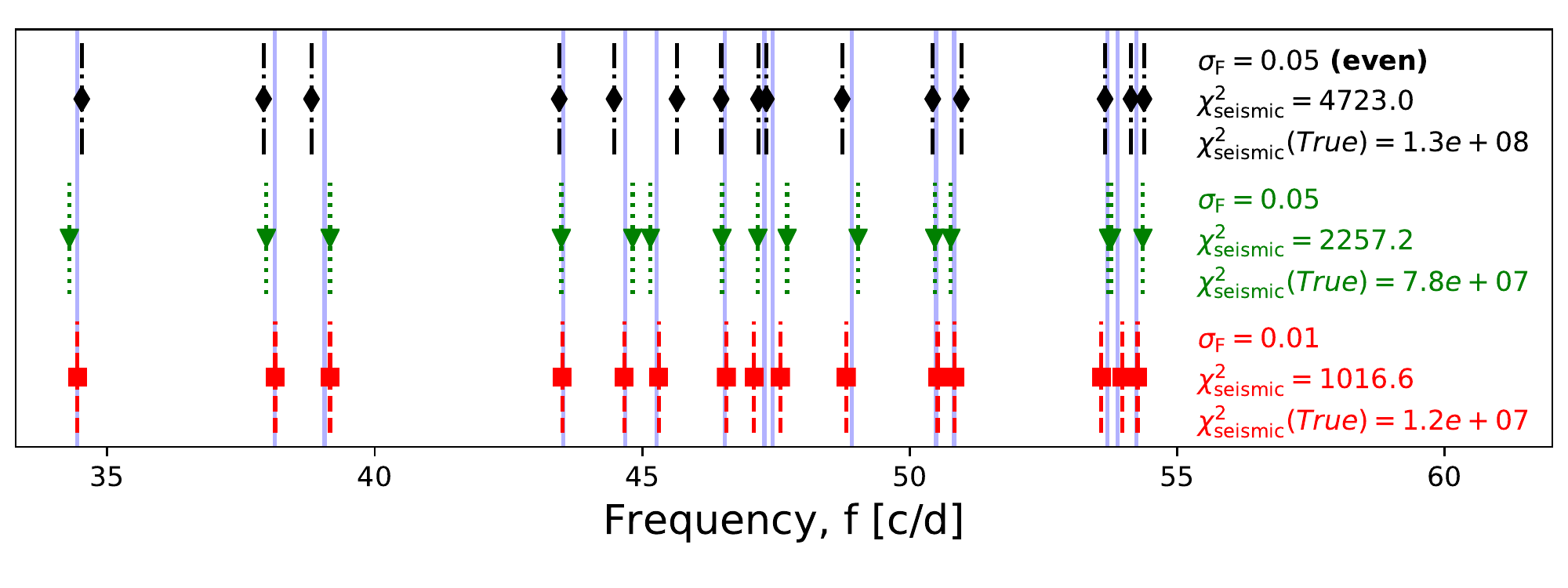}
\caption{(colour online) Same as Fig.~\ref{fig:amplitudes_main} except that $\sigmaF=0.01$ (all modes and inclinations), $\sigmaF=0.05$ (all modes and inclinations), and $\sigmaF=0.05$ (even modes and inclination between $70^{\circ}$ and $90^{\circ}$) have been used during the MCMC runs.  The $\chi_{\mathrm{seismic}}^2$ values are calculated using $\sigmaF=0.01$ regardless of the values $\sigmaF$ used during the MCMC runs, in order to allow direct comparison. Upper panel: For the case of  $\sigmaF=0.05$ using only even modes, $\chi^2_{ampl}$ is 449.1, for the case of $\sigmaF=0.05$ the $\chi^2_{ampl}$ value is 412.7 and for the case of $\sigmaF=0.01$ the $\chi^2_{ampl}$ value is 468.8. }
\label{fig:amplitudes_appendix}
\end{center}
\end{figure*}

Figure~\ref{fig:amplitudes_appendix} shows the best matches from the MCMC runs between observed and theoretical multi-colour amplitudes and spectra using $\sigmaF=0.01$ and $\sigmaF=0.05$ \cd\ as tolerances on the frequencies and restricting the theoretical spectrum to even modes and the inclination to the $[70^{\circ}, 90^{\circ}]$ interval in the last case.

Table C.1. lists the complete theoretical model frequency spectra from our best fitting solutions. The frequencies in bold face
are those which matched $\beta$ Pic's pulsations. Those in italics were filtered out before
the fitting process because they correspond to modes which are antisymmetric with respect
to the equator, and only near equator-on solutions were being searched for in those
particular MCMC runs. We note that these predicted frequencies are subject to large uncertainties in the models, and will change significantly if the input physics will be improved. 

\clearpage

\includepdf[pages=-]{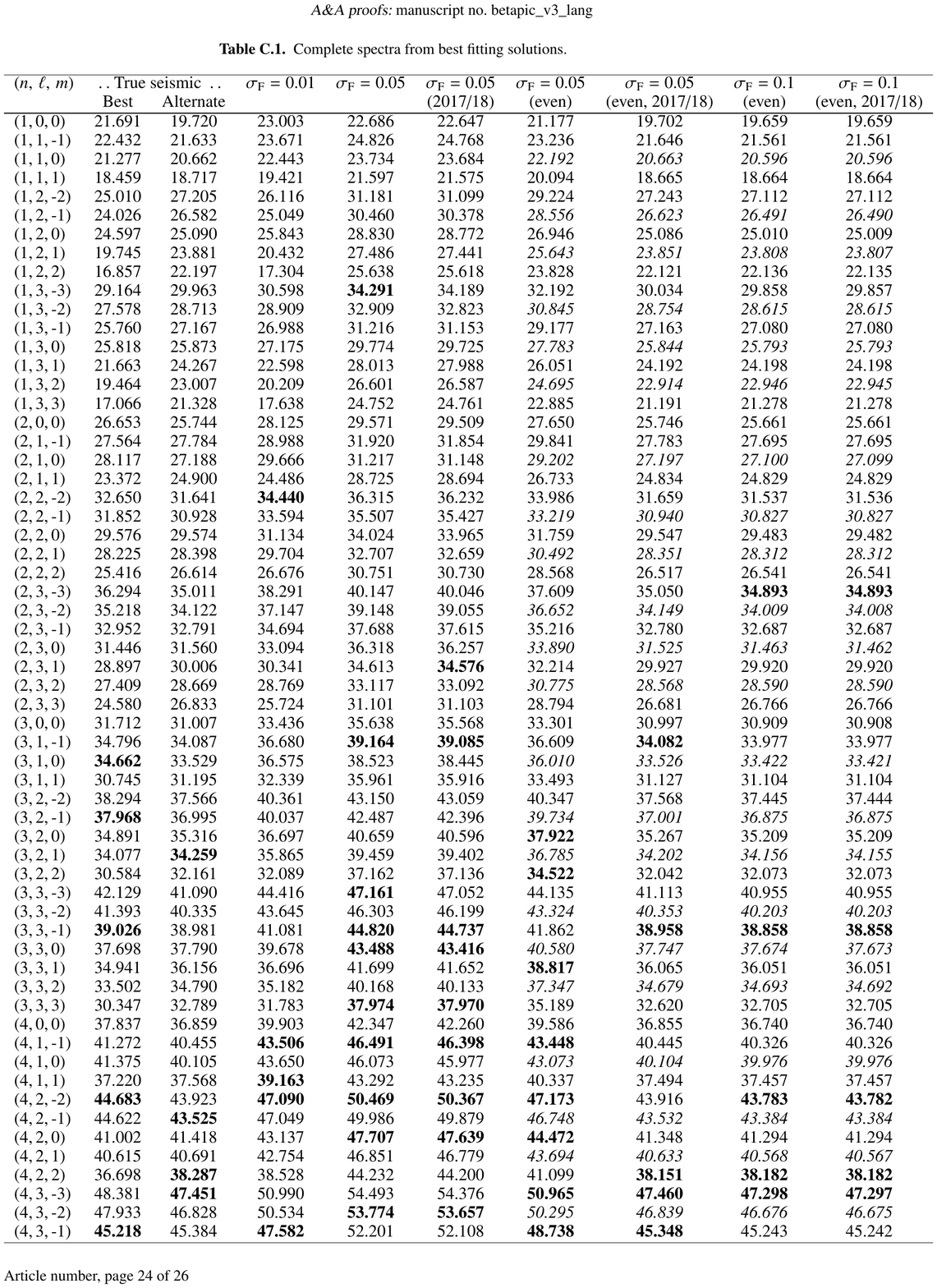}

\end{document}